\documentclass{elsart}

\usepackage{amsmath,amssymb,graphicx,bm}

\newcommand{\beq}{\begin{equation}}
\newcommand{\eeq}{\end{equation}}
\newcommand{\bea}{\begin{eqnarray}}
\newcommand{\eea}{\end{eqnarray}}
\newcommand{\ben}{\begin{eqnarray*}}
\newcommand{\een}{\end{eqnarray*}}

\newcommand{\vlowk}{V_{{\rm low}\,k}}

\newcommand{\openone}{\leavevmode\hbox{\small1\normalsize\kern-.33em1}}

\newcommand{\Hzero}{T_{D}}
\newcommand{\flow}{s}

\newcommand{\infm}{\,\mbox{fm}^{-1}}

\newcommand{\hbaromega}{$\hbar\Omega$}
\newcommand{\NthreeLO}{N$^3$LO}
\newcommand{\Nmax}{N_{\rm max}}

\newcommand{\Hefour}{$^4$He}
\newcommand{\Hesix}{$^6$He}

\newcommand{\Lisix}{$^6$Li}
\newcommand{\Liseven}{$^7$Li}

\begin{document}


\begin{frontmatter}

\title{Convergence in the no-core shell model \\
with low-momentum two-nucleon interactions}

\author[OSU,MSU]{S.K.\ Bogner},
\ead{bogner@nscl.msu.edu}
\author[OSU]{R.J.\ Furnstahl\corauthref{cor}},
\corauth[cor]{Corresponding author.}
\ead{furnstahl.1@osu.edu}
\author[ISU]{P.\ Maris},
\ead{pmaris@iastate.edu}
\author[OSU]{R.J.\ Perry},
\ead{perry@mps.ohio-state.edu}
\author[TRIUMF]{A.\ Schwenk},
\ead{schwenk@triumf.ca}
\author[ISU]{J.P.\ Vary}
\ead{jvary@iastate.edu}
\address[OSU]{Department of Physics, The Ohio State University, Columbus, OH 43210}
\address[MSU]{National Superconducting Cyclotron Laboratory and Department 
  of Physics and Astronomy, Michigan State University, East Lansing, MI 48844}
\address[ISU]{Department of Physics and Astronomy, Iowa State University, 
              Ames, IA\ 50011}
\address[TRIUMF]{TRIUMF, 4004 Wesbrook Mall, Vancouver, BC, Canada, V6T 2A3}

\date{\today}

\begin{abstract}
The convergence of no-core shell model (NCSM) calculations using
renormalization group evolved low-momentum two-nucleon interactions is
studied for light nuclei up to $^7$Li. Because no additional
transformation was used in applying the NCSM framework, the energy
calculations satisfy the variational principle for a given Hamiltonian.
Dramatic improvements in convergence are found as the cutoffs are
lowered.  
The renormalization group equations are truncated at two-body
interactions, so the evolution is only approximately unitary and  
converged energies for $A \geqslant 3$ vary with the cutoff.  This
approximation is systematic, however, and for useful cutoff ranges
the energy variation
is comparable to natural-size truncation errors inherent from the
initial chiral effective field theory potential.
\end{abstract}

\end{frontmatter}
\maketitle


\section{Introduction}
\label{sec:intro}

The nonperturbative nature of inter-nucleon interactions is strongly
scale or resolution  dependent and can be radically altered by using
renormalization group (RG) methods 
to decouple low- and high-momentum modes in nuclear
Hamiltonians~\cite{VlowkRG,Vlowk1,Vlowk2,BognerNucmatt,%
Bogner:2006vp,Bogner:2006srg,Hergert:2007wp}.  
A consequence of this decoupling is that few- and many-body
calculations become more tractable at lower 
resolutions~\cite{Vlowk3N,Bogner:2006tw,Variational,Hagen:2007ew,Hagen:2007hi}, 
which
implies that {\it ab-initio} basis expansion methods such as the No-Core
Shell Model (NCSM) should exhibit much improved convergence
properties. 
Moreover, the RG approach has the  advantage of
being able to vary the cutoff as a tool to probe the
quality of the many-body approximation and to provide estimates of the size
of omitted terms in the Hamiltonian 
(e.g., three-nucleon interactions)~\cite{Vlowk3N}.  

In this paper, we present NCSM calculations of
ground states and some low-lying natural-parity excited states 
for light nuclei 
($^3$H, $^4$He, $^6$He, $^6$Li and $^7$Li) using RG-evolved
nucleon-nucleon (NN) interactions as input.
These potentials include similarity renormalization group (SRG) interactions
and both sharp~\cite{Vlowk2} and
smooth~\cite{Bogner:2006vp} 
cutoff versions of low-momentum interactions $\vlowk$ 
derived from one of the N$^3$LO effective field theory (EFT) two-nucleon
potentials~\cite{N3LO,N3LOEGM}. 
The dependence of convergence properties on the cutoff
is of particular interest.%
\footnote{The SRG potentials are characterized by a flow parameter $\lambda$
that sets the scale of decoupling of low-energy and high-energy physics,
thus playing the role of a cutoff.  For convenience 
we will refer to both the SRG $\lambda$ and the $\vlowk$ $\Lambda$ as cutoffs.}
In applying the NCSM framework,
no additional transformation of the Hamiltonian is performed.
Thus, the energy calculations here satisfy the variational principle 
for the NN interaction at each cutoff.
Details of the NCSM approach can be found in 
Refs.~\cite{NCSMC12,NCSM2,NCSM3,Nogga:2005hp,Navratil:2007we}
and previous applications of the 
NCSM to soft bare NN potentials are described in
Refs.~\cite{Shirokov:2003kk,Shirokov:2004ff,Shirokov:2005bk}.

All of the calculations are for NN interactions only.
Since the physics is qualitatively
incomplete without including at least three-nucleon (3N) interactions,
these should be viewed as baseline calculations that set the stage for
inclusion of 3N interactions in the near future~\cite{SRG3body}; some
areas of investigation have been deferred until then.
The additional computational costs of 3N interactions
will severely limit the feasible basis sizes for all but the lightest
nuclei.
So rather than use large spaces to ensure convergence of all the NN-only
calculations,
we have used relatively small spaces for all nuclei. 
With the improved convergence rates at lower
momentum cutoffs, we have been able to perform all of the
calculations on a small computer cluster.  
This is sufficient to establish that the scaling of computer resources
should be favorable for adding three-body interactions.

The RG equations inevitably shift strength from two-body to many-body
interactions as $\lambda$ or $\Lambda$ 
is lowered, even if many-body interactions are initially zero. 
If we only keep the NN part, the transformations are only approximately 
unitary and
few-body binding energies will change  (``run'')
with $\lambda$ or $\Lambda$.
This approximation is not a problem because the initial chiral EFT
potentials have truncation errors and associated theoretical error
bands for observables.  We can
ensure that the variation of energies (and other observables) 
are comparable to the EFT truncation errors and that the hierarchy
of many-body forces is preserved by restricting the range of RG cutoffs.
By tracking the running of the energies, 
we can assess the expected net 
contributions from many-body forces in a more complete future calculation. 

We stress at the outset that although the converged (or extrapolated)
binding energies for certain cutoffs are quite close to experimental
binding energies, this should not be taken as motivation for
using those cutoffs and neglecting three-body interactions.
Since nuclear matter calculated with low-momentum interactions
does not saturate with NN only~\cite{BognerNucmatt}, it is clear that 
the importance of 3N contributions will increase as one
moves higher in the table of nuclides.

The outline of the paper is as follows.
In Section~\ref{sec:srgenergies} we look at the convergence of 
ground-state energies of light nuclei in a harmonic-oscillator basis
using SRG-evolved momentum-space interactions.  
We find that decoupling documented for two-body systems in
Ref.~\cite{Bogner:2007srg} is also present in heavier nuclei (with NN only).
The SRG convergence patterns are compared to those for $\vlowk$ potentials
using regulators with varying degrees of smoothness in
Section~\ref{sec:vlowkenergies}. 
A simple extrapolation procedure, which is
needed for the larger cutoffs in nuclei
with $A>3$ and for all cutoffs in $^7$Li, yields energies with error
bars that exhibit the net evolution (``running'') of three-body 
(and higher-body)
contributions.  
The error bars reflect the NCSM convergence and the cutoff variation
assesses the impact of missing many-body interactions.
These results are exhibited in Section~\ref{sec:running}.
Convergence for selected excited states and radii are shown in
Section~\ref{sec:observables}.
We conclude in Section~\ref{sec:summary} 
with a summary and outlook for future calculations.

 
\section{Convergence of SRG Ground-State Energies}
\label{sec:srgenergies}

In this section, we present results for ground-state energies of light
nuclei for a range of
harmonic-oscillator basis sizes and oscillator parameters using
potentials evolved with the similarity renormalization
group~\cite{Glazek:1993rc,Wegner:1994,Szpigel:2000xj}.
The SRG flow is a series of unitary transformations parametrized
by a flow variable $s$:
\beq
H_\flow = U(\flow) H U^\dagger(\flow) \equiv T + V_\flow \,,
\label{eq:Hflow}
\eeq
where $H$ is the original Hamiltonian (corresponding to $s=0$) and
$T$ is the kinetic energy, which is taken to be independent of $s$
(thereby defining $V_\flow$).
The choice of transformations leads to the flow equation for $V_\flow$,
\beq
\frac{dV_\flow}{d\flow} = [ [\Hzero, H_\flow], H_\flow] 
= [ [\Hzero, V_\flow], H_\flow] \,,
\label{eq:commutator}
\eeq 
which is evaluated here for NN interactions in a partial-wave momentum 
basis~\cite{Bogner:2006srg,Bogner:2007srg,srgwebsite}.  
The operator $\Hzero$ is a diagonal matrix in this basis.  The results
of Refs.~\cite{Bogner:2006srg,Bogner:2007srg} and all results here use
$\Hzero = T$, but we have also checked convergence for $\Hzero = T^2$
and $\Hzero = T^3$.
For $\Hzero = T$ in momentum representation, 
$\lambda \equiv 1/s^{1/4}$ is a more useful flow variable
that can be thought of as a cutoff on momentum transfers, and
we use it exclusively from now on.

We use primarily
the N$^3$LO potential from Ref.~\cite{N3LO}, which has been
the subject of previous investigations in the
NCSM~\cite{Nogga:2005hp,Navratil:2007we}.  In those investigations, 
Lee-Suzuki
transformations were used to derive effective interactions that 
improved convergence, but which lose the variational property when making
a cluster approximation.  In contrast,
the present calculations use interactions evolved in free space, so they simply
represent alternative Hamiltonians; thus the calculations retain
the variational principle.
Because the evolution to lower momenta yields nearly universal interactions
for $\lambda \lesssim 2\infm$~\cite{VlowkRG,Vlowk2,Bogner:2006vp,Bogner:2006srg}, 
convergence properties starting from
other initial N$^3$LO~\cite{N3LOEGM} 
or phenomenological potentials~\cite{AV18} 
are quantitatively similar.
For simplicity, in these first NCSM calculations we evolve only the  
neutron-proton part of nuclear forces, with Coulomb added after the
RG evolution.

The choice of $\Hzero = T$ in Eq.~(\ref{eq:commutator}) 
leads to a partial diagonalization of
the potentials in momentum representation, which is the
source of decoupling of low- and high-momentum contributions.  We expect
reduced short-range correlations and
improved convergence  in almost any basis as a result of this decoupling.
The evolution modifies only the short-distance part of operators;
for the Hamiltonian
this is important to maintain the hierarchy of many-body forces
in the initial EFT Hamiltonian (as long as we don't evolve too far).
However, one could also consider tailoring the evolution to a particular
basis such as harmonic oscillators. 
In this case, choosing $\Hzero$ to be the one-body harmonic oscillator
Hamiltonian (with a two-body term added to the potential as usual
with the NCSM~\cite{NCSMC12,NCSM2,NCSM3}) will yield a
flow to more diagonal and therefore more decoupled potentials in this basis,
accelerating convergence. 
The downside is that there may be serious negative consequences for
long-distance operators and many-body forces,
analogous to what is observed for Lee-Suzuki transformations in the
NCSM~\cite{Stetcu:2004wh}.
These issues will be explored in a future investigation.

Matrix elements of the momentum-space potential are evaluated in
a harmonic-oscillator basis for input to the Many-Fermion Dynamics (MFD)
code~\cite{MFD}, which constructs the many-body matrix elements and
performs the diagonalization.
The basis is specified by the
oscillator parameter \hbaromega\ and by $\Nmax$.
$\Nmax$ defines the maximum number of oscillator 
quanta (increments based on sums of single 
particle quanta, $2n + l$) allowed above the 
lowest many-body configuration for a given 
nucleus.  Thus $\Nmax = 2$ implies the inclusion of 
single-particle excitations up to two major 
oscillator shells or two particles excited 
simultaneously by one major shell above the 
lowest-energy oscillator configuration.

We view the present NN-only calculations as setting the baseline
for upcoming calculations that will include 3N forces.
Because the latter will have significant computational costs
that will limit the maximum basis size that can be used,
we study convergence for different $\lambda$ values with
at most $\Nmax = 12$ and
dimension sizes of order $10^7$ or less.
This enables a significant number of exploratory cases to be 
easily and quickly examined on a small computer cluster, 
including the evaluation of a suite of 
experimental observables for each case.
The basis dimension for $^2$H ranges from 24 for $\Nmax = 2$ to 4200 for
$\Nmax = 12$, for $^6$Li from 800 for $\Nmax = 2$ to 9,692,634
for $\Nmax = 10$ and  for $^7$Li from 1961 
for $\Nmax = 2$ to 6,150,449 for $\Nmax = 8$.
Extending these calculations to more powerful computers for 
including 3N forces is straightforward.
   
\begin{figure}[pth]
  \centerline{\includegraphics*[width=4.85in,angle=0]%
             {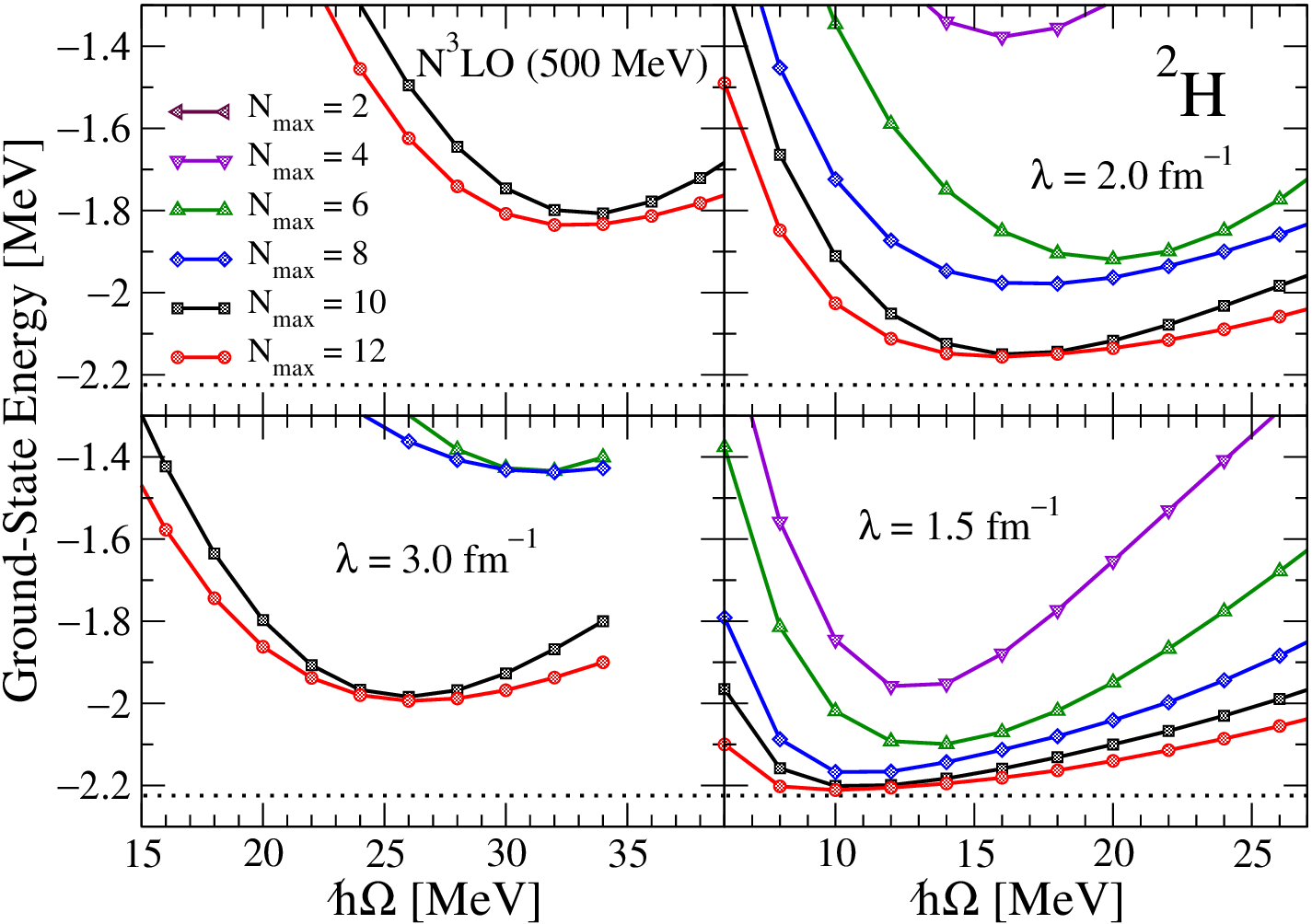}}
  \vspace*{-.1in}
  \caption{Ground-state energy of the deuteron as a function of \hbaromega\
    at four different values of $\lambda$ ($\infty$, 3, 2, $1.5\infm$)
    for $\Nmax$ values up to 12.
            The initial potential is the 500\,MeV \NthreeLO\ NN-only
	    potential from Ref.~\cite{N3LO}. The dotted
	    lines show fully converged energies obtained with
	    a sufficiently large basis size.}
  \label{fig:H2_srg_composite}

  \vspace*{.05in}

  \centerline{\includegraphics*[width=4.85in,angle=0]%
             {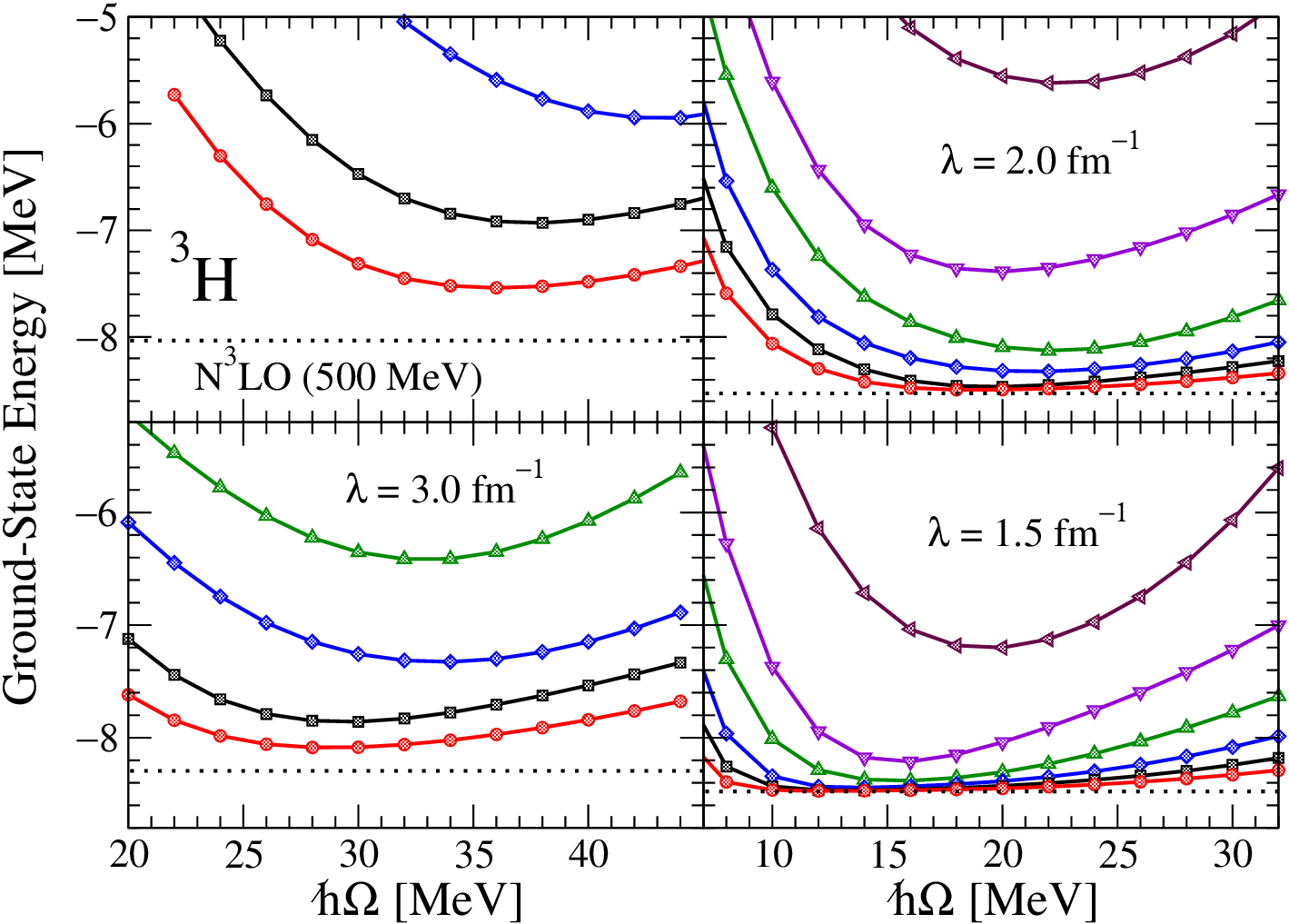}}
  \vspace*{-.1in}
  \caption{Ground-state energy of the triton as a function of \hbaromega\
    at four different values of $\lambda$ ($\infty$, 3, 2, $1.5\infm$).
            The initial potential is the 500\,MeV \NthreeLO\ NN-only
	    potential from Ref.~\cite{N3LO}. The legend from
            Fig.~\ref{fig:H2_srg_composite} applies here.  The dotted
	    lines show fully converged energies from independent
	    $\Nmax = 48$  calculations using a code from
	    Ref.~\cite{Navratil00}.}
  \label{fig:H3_srg_composite}
\end{figure}

\begin{figure}[pth]
  \centerline{\includegraphics*[width=4.85in,angle=0]%
             {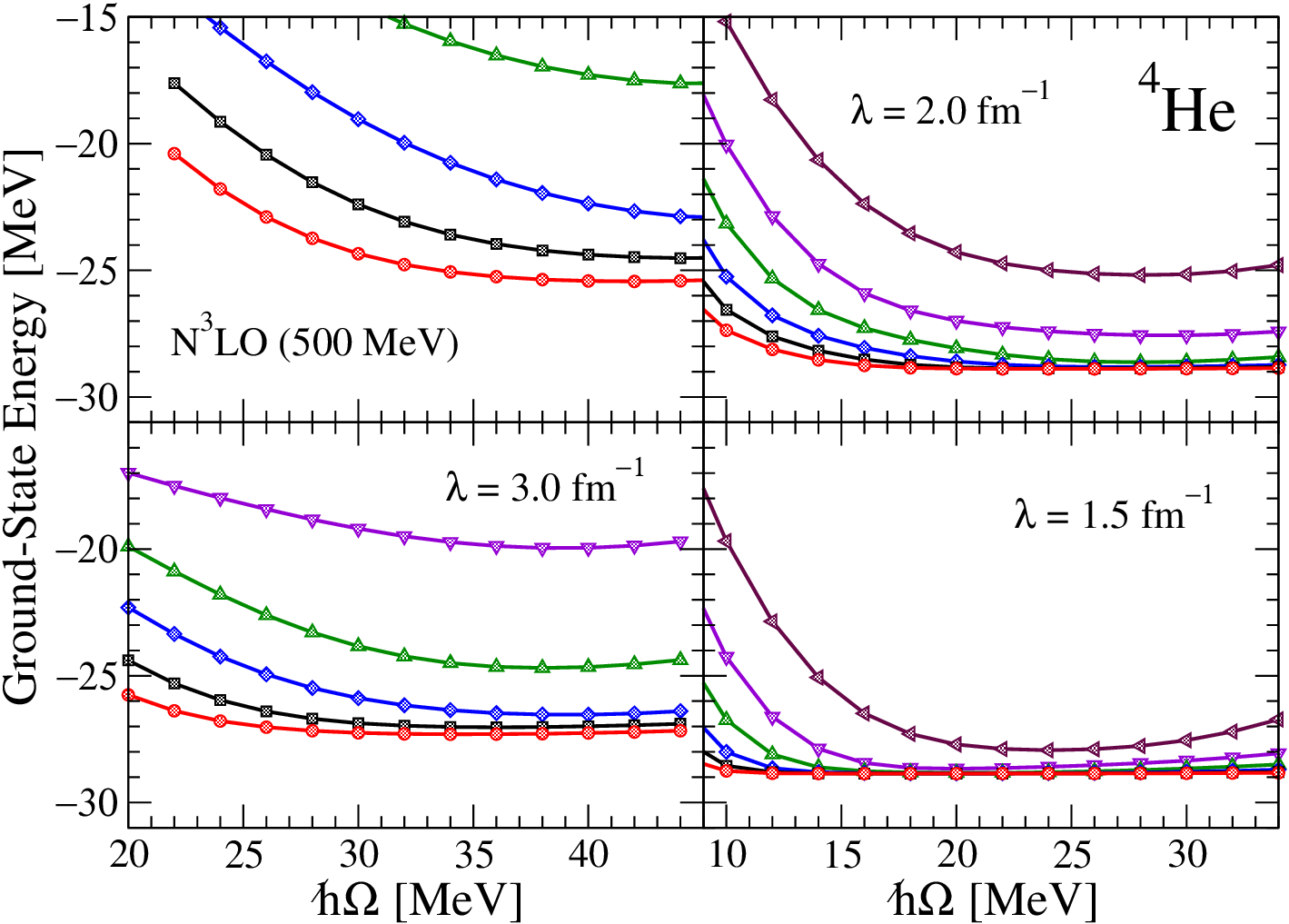}}
  \vspace*{-.1in}
  \caption{Ground-state energy of \Hefour\ as a function of \hbaromega\
    at four different values of $\lambda$ ($\infty$, 3, 2, $1.5\infm$).
            The initial potential is the 500\,MeV \NthreeLO\ NN-only
	    potential from Ref.~\cite{N3LO}.
            The legend from
            Fig.~\ref{fig:H2_srg_composite} applies here.}
  \label{fig:He4_srg_composite}
  \vspace*{.1in}

  \centerline{\includegraphics*[width=4.85in,angle=0]%
             {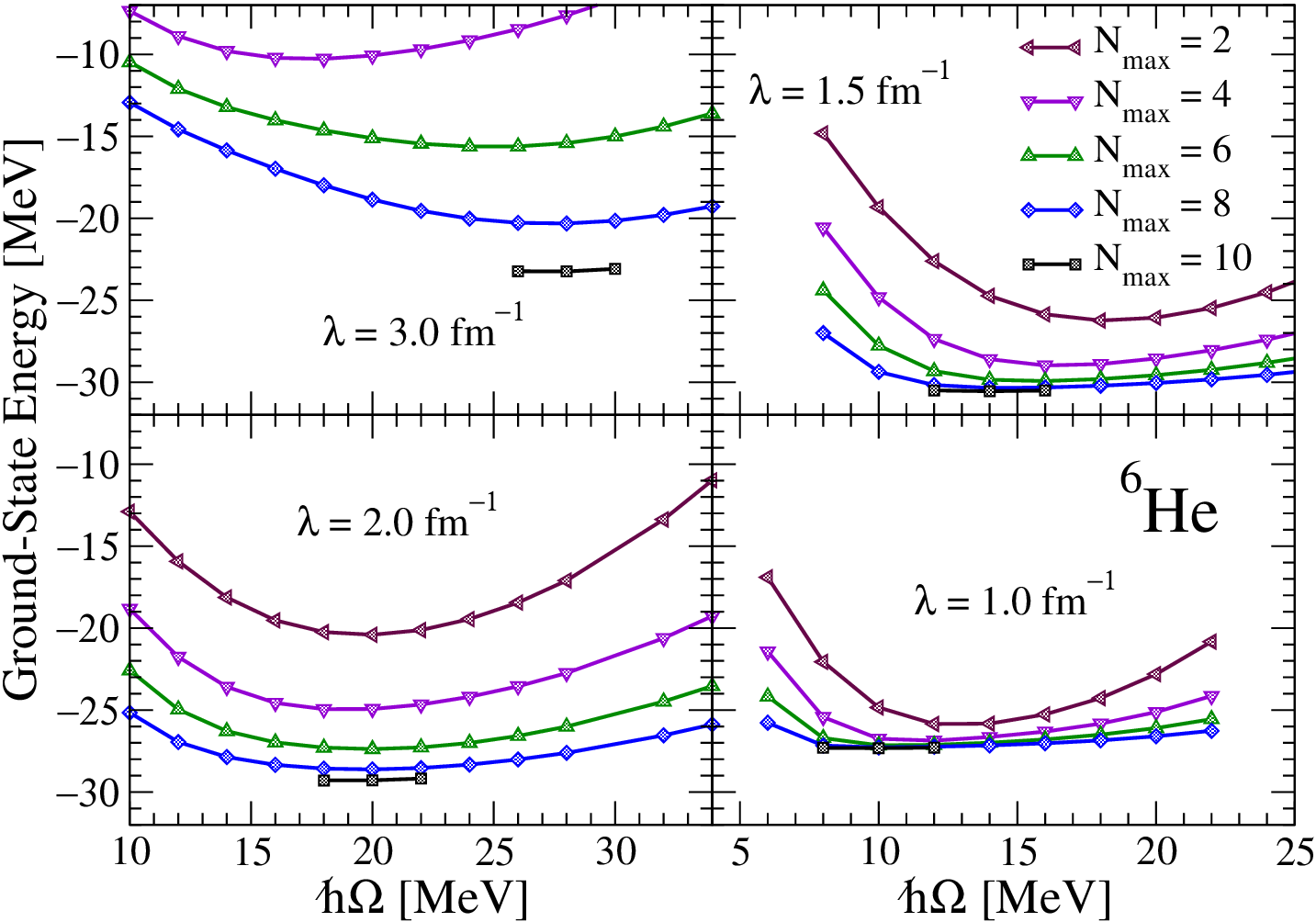}}
  \vspace*{-.1in}
  \caption{Ground-state energy of \Hesix\ as a function of \hbaromega\
    at four different values of $\lambda$ (3, 2, 1.5, $1\infm$).
            The initial potential is the 500\,MeV \NthreeLO\ NN-only
	    potential from Ref.~\cite{N3LO}.  }
  \label{fig:He6_srg_composite}
\end{figure}

\begin{figure}[pth]
  \centerline{\includegraphics*[width=4.85in,angle=0]%
             {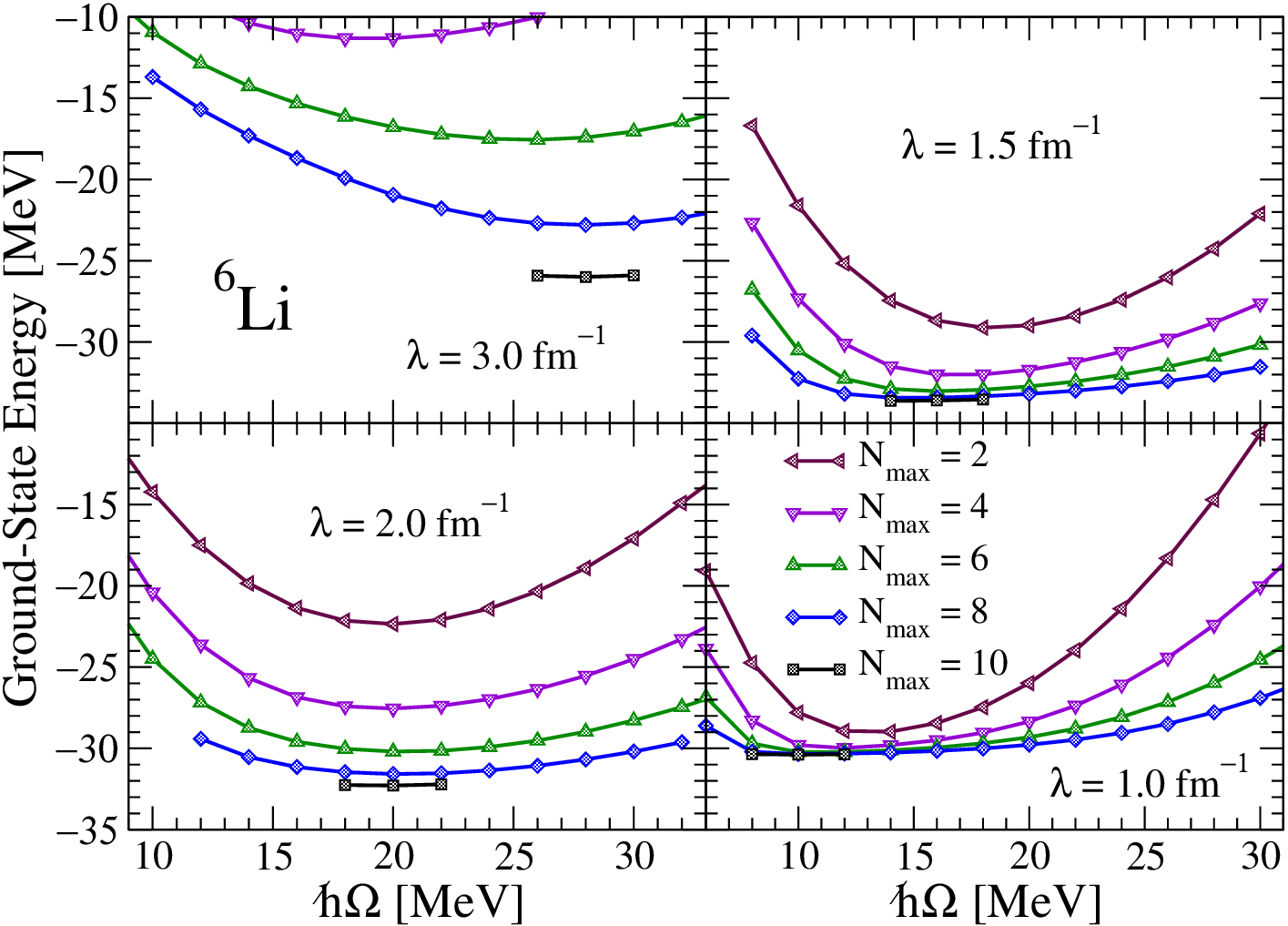}}
  \vspace*{-.1in}
  \caption{Ground-state energy of \Lisix\ as a function of \hbaromega\
    at four different values of $\lambda$ (3, 2, 1.5, $1\infm$).
            The initial potential is the 500\,MeV \NthreeLO\ NN-only
	    potential from Ref.~\cite{N3LO}.}
  \label{fig:Li6_srg_composite}
  \vspace*{.1in}

  \centerline{\includegraphics*[width=4.85in,angle=0]%
            {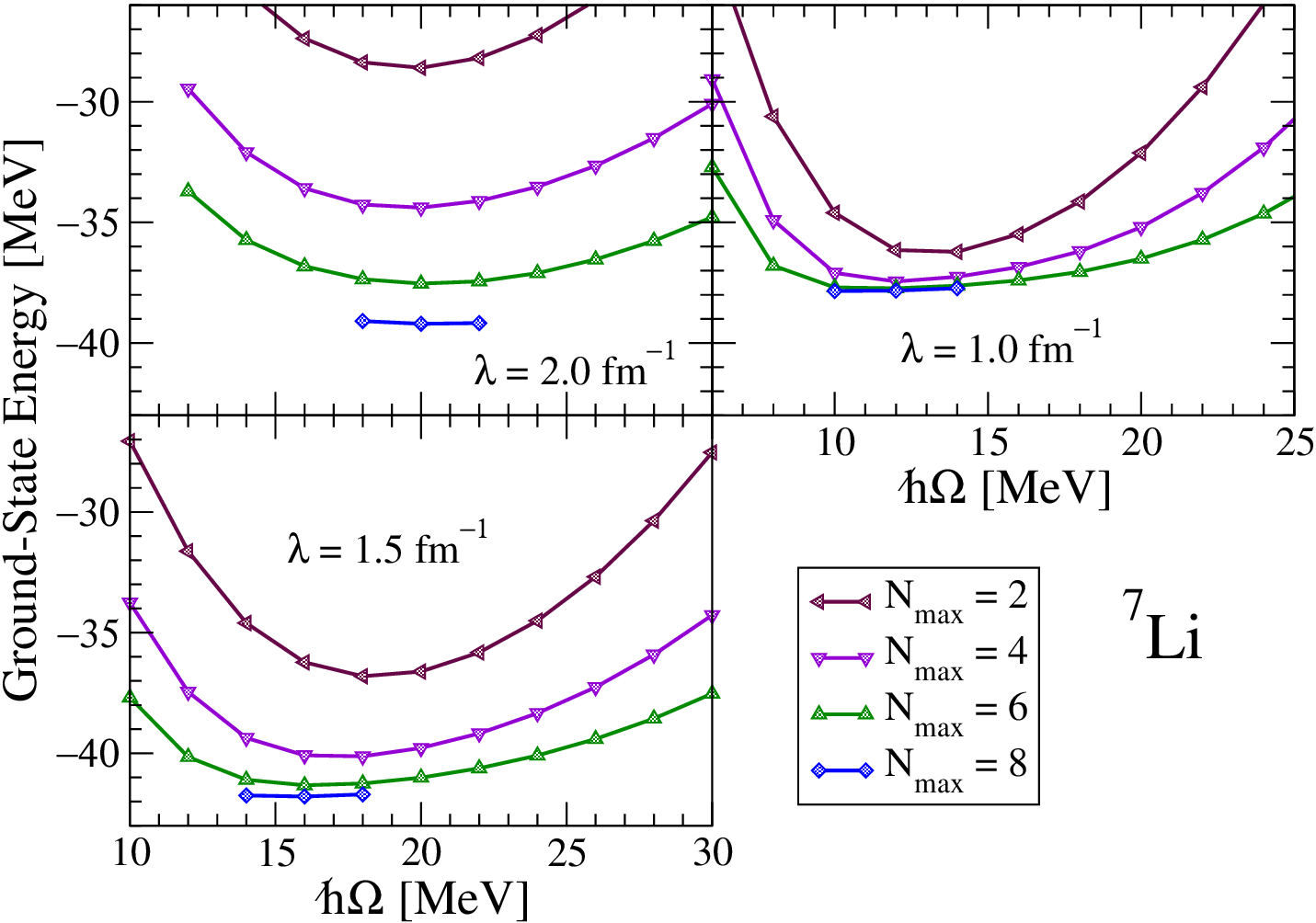}}
  \vspace*{-.1in}
  \caption{Ground-state energy of \Liseven\ as a function of \hbaromega\
    at three different values of $\lambda$ ($2$, $1.5$, $1\infm$).
            The initial potential is the 500\,MeV \NthreeLO\ NN-only
	    potential from Ref.~\cite{N3LO}.  }
  \label{fig:Li7_srg_composite}
\end{figure}

In Figs.~\ref{fig:H2_srg_composite} through
\ref{fig:Li7_srg_composite},
the ground-state energies of
$^2$H, $^3$H, $^4$He, $^6$He, $^6$Li, and $^7$Li
are plotted against the harmonic-oscillator
parameter \hbaromega\ for a range of $\Nmax$ values.  
The figures with $A \leqslant 4$ show results for the initial
chiral N$^3$LO potential and three SRG-evolved potentials with
$\lambda = 3.0$, $2.0$, and $1.5\infm$.
For $A=6$ and $A=7$, our largest spaces were very far from
convergence with the initial
potential, so we show only SRG-evolved potentials
with $\lambda = 3.0$, $2.0$, $1.5$, and $1.0\infm$.
 
It may appear in Fig.~\ref{fig:H2_srg_composite} 
that the deuteron is converged to different 
ground-state energies for different $\lambda$ values.
However, the results simply reflect the slow convergence
of the weakly bound deuteron in a harmonic oscillator basis;
all converge with sufficiently large spaces
to the same energy as the unevolved potential.
Only for $\lambda = 1.5\infm$ is the energy   
converged at the 10\,keV level by $\Nmax = 12$.
(For the other cutoffs, 
there is an irregular convergence trend for which
$\Nmax=12$
is very close to $\Nmax = 10$, but then $\Nmax = 14$ will be lowered
further.)  
The convergence rate for $^3$H in Fig.~\ref{fig:H3_srg_composite} 
is similar when viewed on the same scale.
In contrast to the deuteron, the convergence is to noticeably
different energies in large spaces (indicated by the dotted lines)
because the SRG evolution truncated to NN-only is only approximately
unitary for $A \geqslant 3$.  The spread of energies is a measure
of this approximation; for the range of $\lambda$'s shown, it is the
same order as the truncation error from omitting 3N forces in the
original Hamiltonian.  The running of energies
and its implications are discussed
further in Section~\ref{sec:running}.  

In all cases we find rapidly improving convergence with lowered $\lambda$
down to $\lambda = 1.0\infm$, which is consistent with an expected  
decrease in correlations in ground-state wave functions.
Power-counting arguments~\cite{Vlowk3N} 
imply that the hierarchy of many-body forces in heavier nuclei could break
down by this point
(i.e., the 3N and higher-body contributions will become comparable
to the NN contribution). 
However, the scaling of many-body forces in the SRG has not yet
been established, so the practical lower limit in $\lambda$ for which
many-body forces are under control is
not yet known.  Also, it may be different for light and heavy nuclei. 
Once we have established the
technology to consistently run many-body forces with the SRG
(Ref.~\cite{SRG3body} is a first step in this direction), 
we will have greater freedom to
exploit the choice of $\lambda$ in practical calculations.  

In general, the curves of energy vs.\ \hbaromega\ and the trends
with increasing $\Nmax$ at fixed \hbaromega\ are more systematic
with lower $\lambda$.
For example, compare $^6$Li at $\lambda = 3.0\infm$ to
$\lambda = 1.5$ or $2.0\infm$. 
It is also clear that the larger nuclei have more regular behavior
at any $\lambda$.
This implies that extrapolations to $\Nmax = \infty$ will be more
robust for heavier nuclei. 

The trends with smaller $\lambda$'s include increasingly reasonable
estimates from calculations with smaller $\Nmax$ values.
For example, $\Nmax = 2(4)$ for $\lambda = 1.5\infm$
is within 1(0.3) MeV of the converged binding energy for $^4$He
and within 5(1.5) MeV for $^6$Li.
To set the scale for how good these estimates are, 
we note that the starting N$^3$LO potential  in $^4$He is at most
bound by 1\,MeV for $\Nmax = 2$  and only 8\,MeV for $\Nmax = 4$.
For $\lambda = 3.0\infm$, $\Nmax = 2(4)$ is still about
14.5(7.3)\,MeV short of the converged binding energy in $^4$He.

At the same time, with smaller $\lambda$'s
the larger $\Nmax$ values converge entirely (at the 10\,keV
level or better) for the lighter
nuclei.  As $\Nmax$ increases,
the  \hbaromega\ 
dependence gets flatter in a  
very smooth and systematic way.
We also note that for a given $\Nmax$,
the minimum in \hbaromega\  moves toward lower values
as $\lambda$ decreases, as is expected for softer interactions.
The energies for $\lambda = 1.0\infm$ converge remarkably fast.
The binding in this case is smaller than for the other $\lambda$
values, which is consistent with the running of  
the net three-body
contribution 
in Ref.~\cite{Vlowk3N} for a sharp $\vlowk$ interaction in
$^3$H and $^4$He.  
Explicit results for the running of the net many-body contribution 
are given in Section~\ref{sec:running}.

We can compare the convergence patterns observed here with those
found in previous NCSM investigations.
The convergence of Lee-Suzuki effective interaction 
results with increasing $\Nmax$ is generally 
non-monotonic at fixed \hbaromega.  This is seen 
in numerous examples in the literature
(see, e.g., Ref.~\cite{Shirokov:2005bk}). It 
is therefore more challenging to make 
extrapolations in those cases.  However, 
extrapolations with bare but soft NN 
interactions, such as JISP16 (Ref.~\cite{Shirokov:2005bk}), are more 
straightforward~\cite{Pieter} and similar to the SRG convergence.
Extrapolations of SRG results are tested in Section~\ref{sec:running}. 

\begin{figure}[pth]
  \centerline{\includegraphics*[width=4.5in,angle=0]%
             {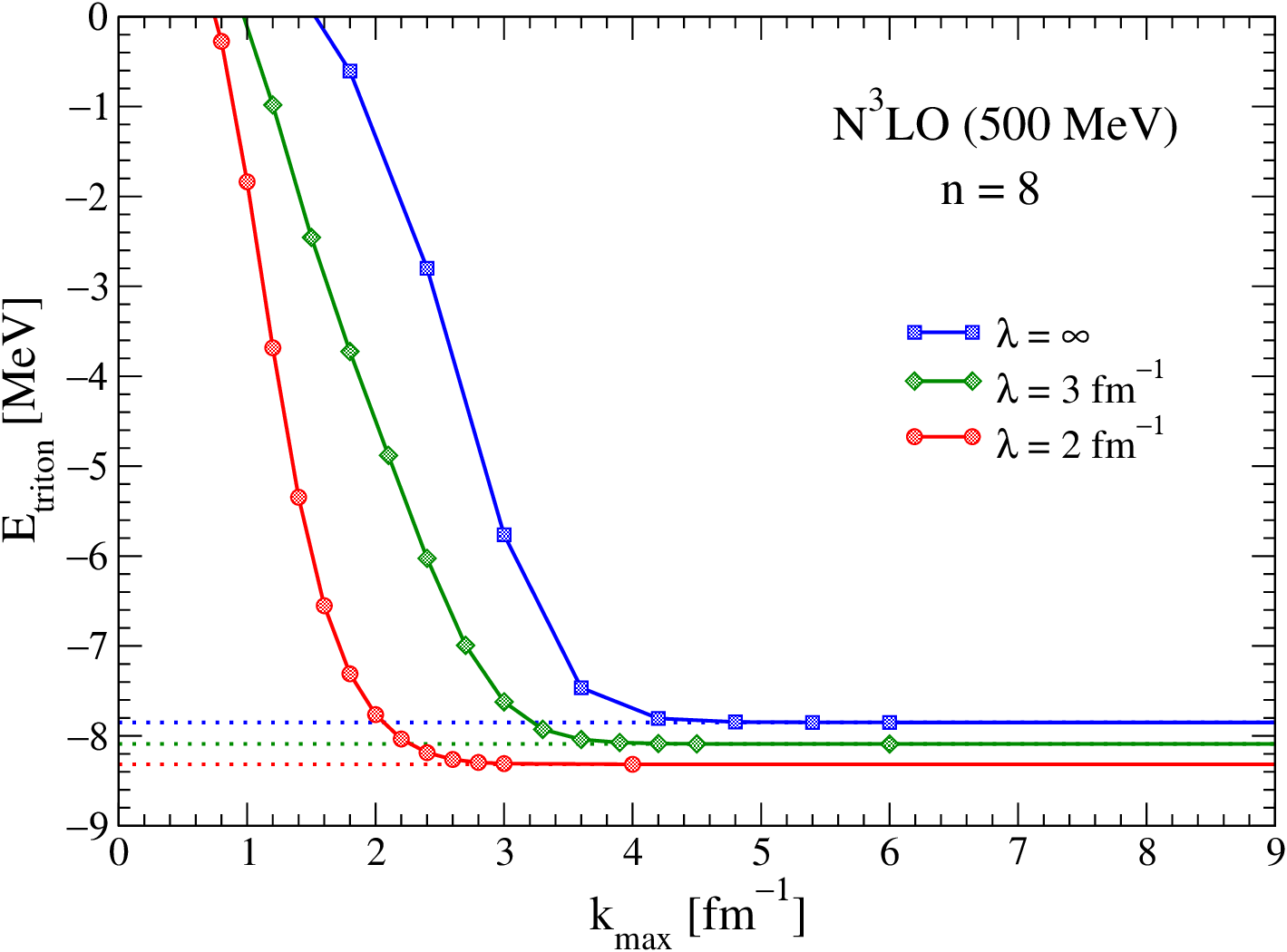}}
  \vspace*{-.1in}
  \caption{Decoupling through the SRG.
   Binding energy as a function of cutoff parameter $k_{\rm max}$
    for three different values of $\lambda$.
   The cutoff function is $\exp[-(k^2/k_{\rm max}^2)^n]$ with $n=8$.
            The initial potential is the 500\,MeV \NthreeLO\ NN-only
	    potential from Ref.~\cite{N3LO}. 
            These results include evolved electromagnetic interactions  
           and charge dependences.
	    }
  \label{fig:H3_srg_decoupling_N3LO_500MeV}

  \vspace*{.15in}

  \centerline{\includegraphics*[width=4.5in,angle=0]%
             {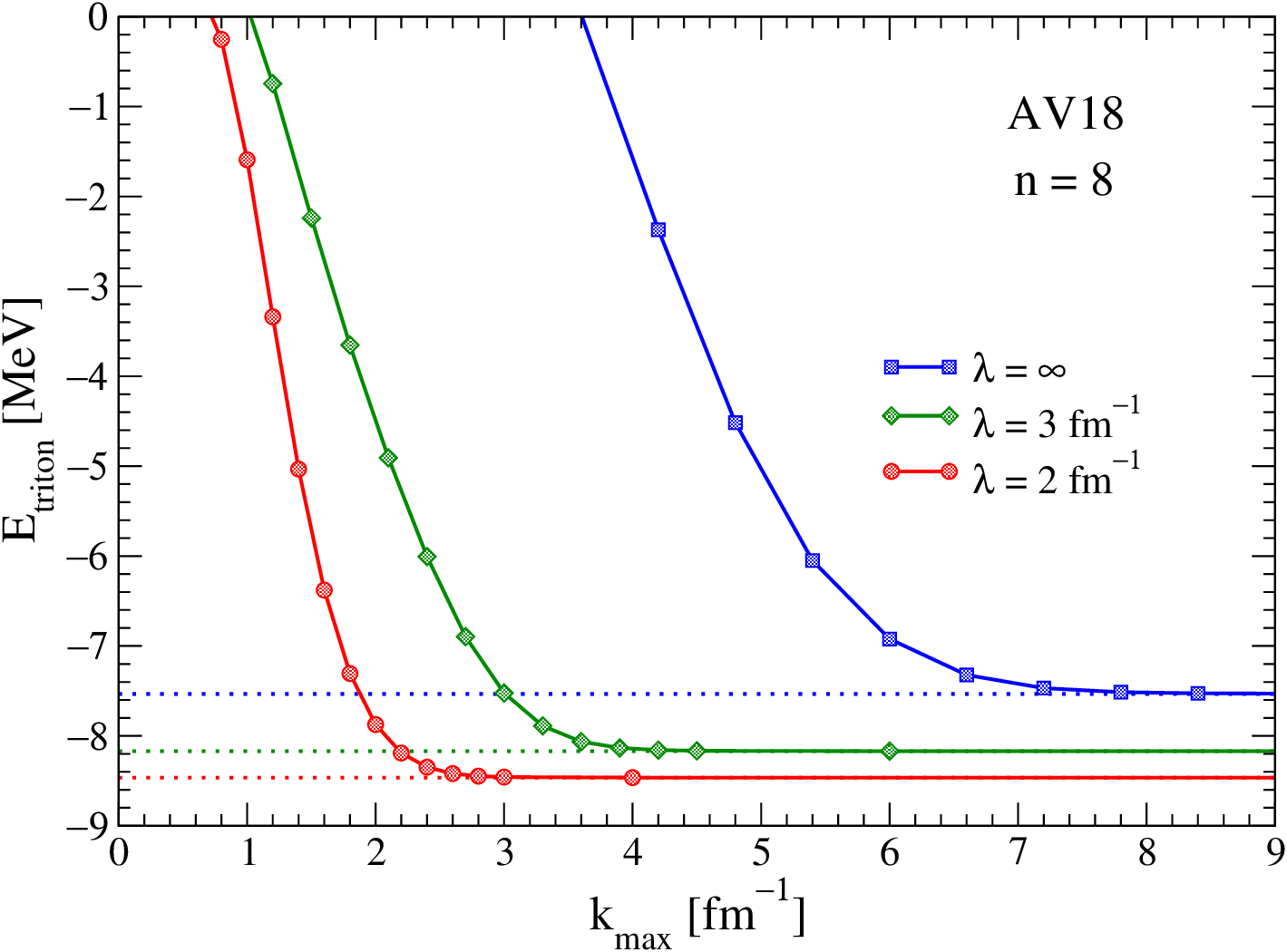}}
  \vspace*{-.1in}
  \caption{Decoupling through the SRG. 
   Ground-state energy as a function of cutoff parameter $k_{\rm max}$
   for three different values of $\lambda$.
   The cutoff function is $\exp[-(k^2/k_{\rm max}^2)^n]$ with $n=8$.
  The initial potential
     is NN-only Argonne $v_{18}$ \cite{AV18}. 
     These results include evolved electromagnetic interactions  
           and charge dependences. }
  \label{fig:H3_srg_decoupling_AV18}
\end{figure}

In Ref.~\cite{Bogner:2007srg}, the SRG was used to demonstrate how
running to lower momentum decouples low- and high-momentum contributions
to matrix elements of low-energy observables.  Only two-body observables
were considered.
Using the NCSM, we can extend the tests of decoupling with NN interactions
to few-body systems.
One test is to apply a cutoff function to the SRG potential, in
the form
\beq
   V_{\lambda}(k,k'; k_{\rm max}) =
     e^{-(k^2/k_{\rm max}^2)^n}  V_{\lambda}(k,k')
     e^{-(k'{}^2/k_{\rm max}^2)^n} \;,
     \label{eq:cutoffV}
\eeq
where the choice for $n$ controls the smoothness of the cutoff, 
and then to calculate the ground-state
energy as $k_{\rm max}$ is varied from 0 to $\infty$.
This second cutoff (here with $n=8$) 
is simply imposed by hand after the RG cutoff is run as a tool
to test decoupling: if there is decoupling of low- and high-momentum
contributions, then we should be able to set matrix elements of the
potential to zero in a smooth way above some momentum $k_{\rm max}$ and still
get the same answer for low-energy observables.
The effect of the second cutoff
is shown for the triton
in Fig.~\ref{fig:H3_srg_decoupling_N3LO_500MeV} starting from the N$^3$LO
potential of Ref.~\cite{N3LO}.  For comparison, we show in  
Fig.~\ref{fig:H3_srg_decoupling_AV18} the same calculation but
starting with the Argonne $v_{18}$ potential \cite{AV18}.

The convergence with increasing $k_{\rm max}$
of the binding energy for the bare ($\lambda=\infty$) interaction
to the asymptotic value is determined by the intrinsic cutoffs 
in the potential and reflects
the associated decoupling.  
Thus for AV18, the energy is converged
to good accuracy
by $k_{\rm max} \approx 7\infm$ while for the N$^3$LO potential
convergence is reached between 4 and $4.5\infm$.  
(Note that the latter result is
much higher than the naive estimate of $2.5\infm$ based on the 500\,MeV
cutoff, due to a significant high-momentum tail and the associated
tensor strength.)
As the potential is evolved to lower $\lambda$, the convergence
scale for the binding energy is set by $\lambda$.  That is, the
most rapid changes happen for $k_{\rm max}$
up to $\lambda$ and then there is
a slower approach to the asymptotic value of the energy,  
which is very well converged in all cases
by about $1.4\, \lambda$.
These details are common for any potential evolved with 
$T$ to define the SRG~\cite{SRGdecoupling}.
Similar results for other nuclei and a quantitative perturbative analysis 
of decoupling will
be given in Ref.~\cite{SRGdecoupling}.


We have also examined the convergence of the ground-state energy
for SRG potentials using $\Hzero = T^2$ and $\Hzero = T^3$ in
Eq.~(\ref{eq:commutator}) rather than the kinetic energy $T$.
The convergence behavior 
in both cases is almost indistinguishable from those
for $T$.

\section{Convergence of $\vlowk$ Ground-State Energies}       
  \label{sec:vlowkenergies}

In this section we compare the convergence for the SRG to that
of $\vlowk$ interactions with sharp and various smooth regulators.
In contrast to the SRG flow equations for $V_\flow$, the $\vlowk$
potentials are based on the invariance of the low-energy NN
T~matrix under changes of a cutoff $\Lambda$ on relative
momenta~\cite{Vlowk1,Vlowk2,Bogner:2006vp}.
This is achieved either through coupled differential equations for the
momentum-space matrix elements of $\vlowk$ or in integral form through
a free-space Lee-Suzuki transformation~\cite{VlowkRG}.
The construction and characteristics of $\vlowk$ interactions with
smooth regulators is described in Ref.~\cite{Bogner:2006vp}.

\begin{figure}[pth]
  \centerline{\includegraphics*[width=4.85in,angle=0]%
             {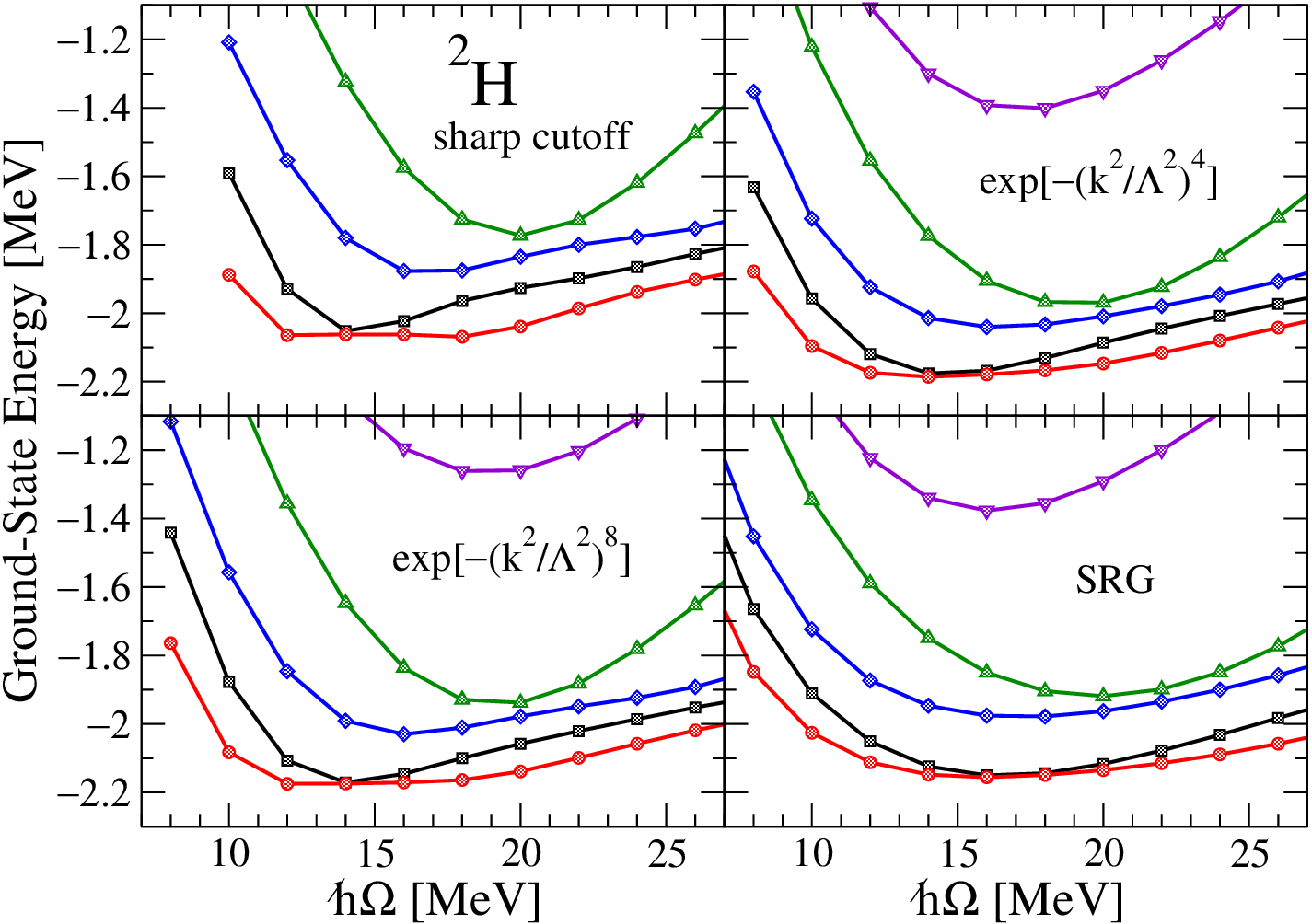}}
  \vspace*{-.1in}
  \caption{Ground-state energy of the deuteron as a function of \hbaromega\
    for a sharp and two smooth $\vlowk$ potentials with $\Lambda = 2\infm$
    and an SRG potential with $\lambda = 2\infm$.
            The initial potential is the 500\,MeV \NthreeLO\ NN-only
	    potential from Ref.~\cite{N3LO}.  The legend from
            Fig.~\ref{fig:H2_srg_composite} applies here.}
  \label{fig:H2_vlowk_composite}
  \vspace*{.1in}

  \centerline{\includegraphics*[width=4.85in,angle=0]%
             {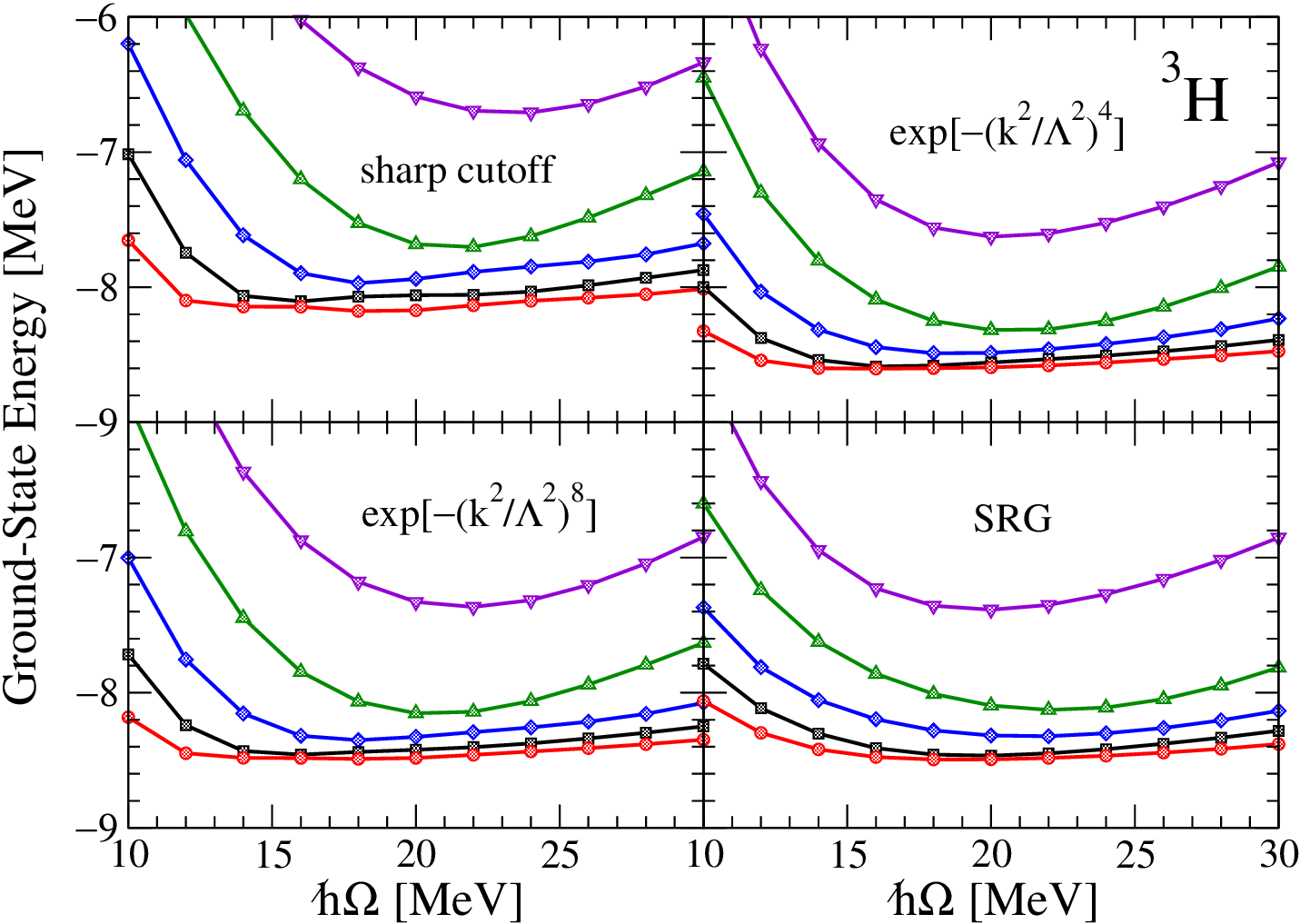}}
  \vspace*{-.1in}
  \caption{Ground-state  energy of the triton\ as a function of \hbaromega\
    for a sharp and two smooth $\vlowk$ potentials with $\Lambda = 2\infm$
    and an SRG potential with $\lambda = 2\infm$.
            The initial potential is the 500\,MeV \NthreeLO\ NN-only
	    potential from Ref.~\cite{N3LO}. The legend from
            Fig.~\ref{fig:H2_srg_composite} applies here.}
  \label{fig:H3_vlowk_composite}
\end{figure}

\begin{figure}[pth]
  \centerline{\includegraphics*[width=4.85in,angle=0]%
             {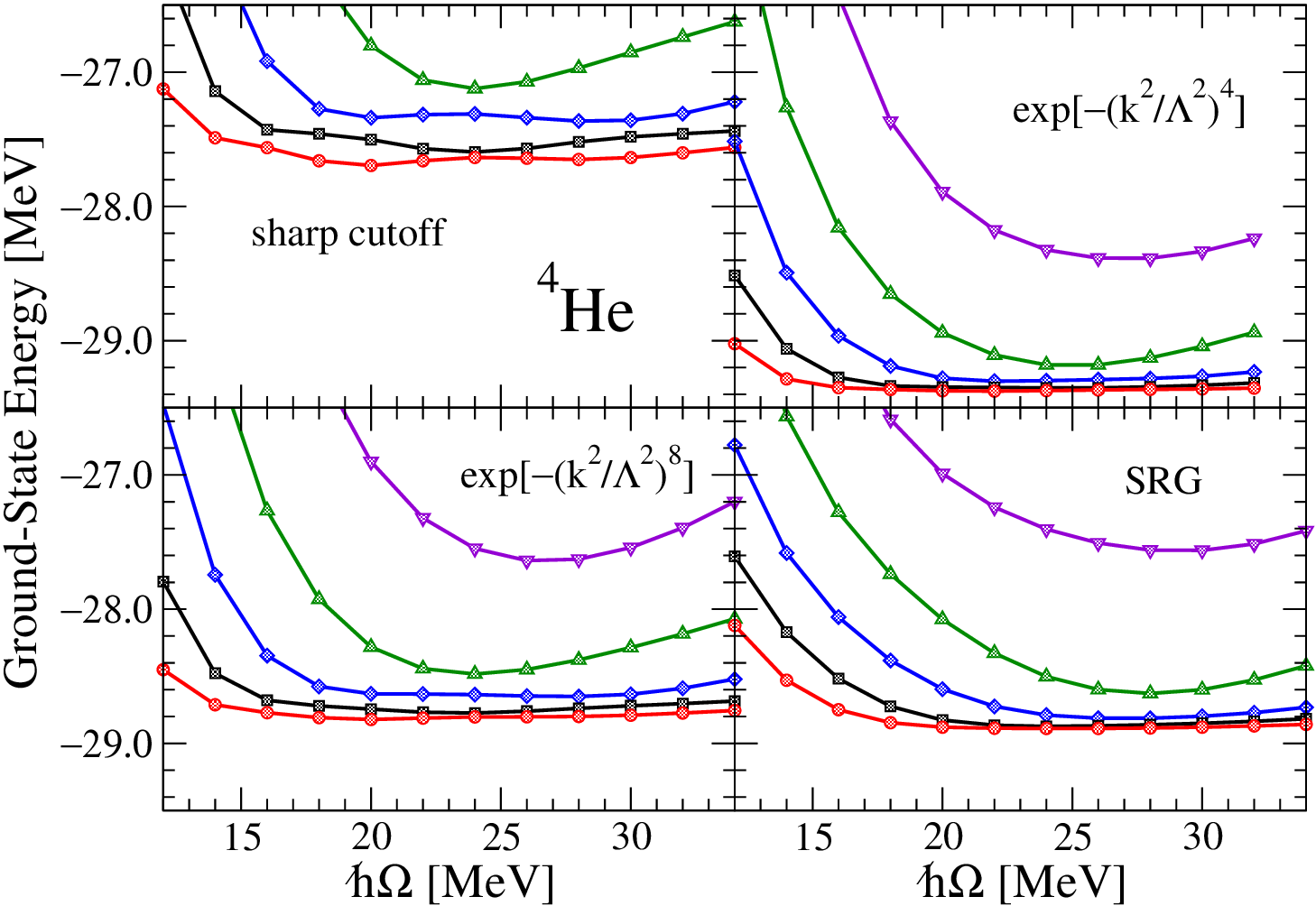}}
  \vspace*{-.1in}
  \caption{Ground-state  energy of \Hefour\ as a function of \hbaromega\
    for a sharp and two smooth $\vlowk$ potentials with $\Lambda = 2\infm$
    and an SRG potential with $\lambda = 2\infm$.
            The initial potential is the 500\,MeV \NthreeLO\ NN-only
	    potential from Ref.~\cite{N3LO}.  The legend from
            Fig.~\ref{fig:H2_srg_composite} applies here.}
  \label{fig:He4_vlowk_composite}
  \vspace*{.1in}

  \centerline{\includegraphics*[width=4.85in,angle=0]%
             {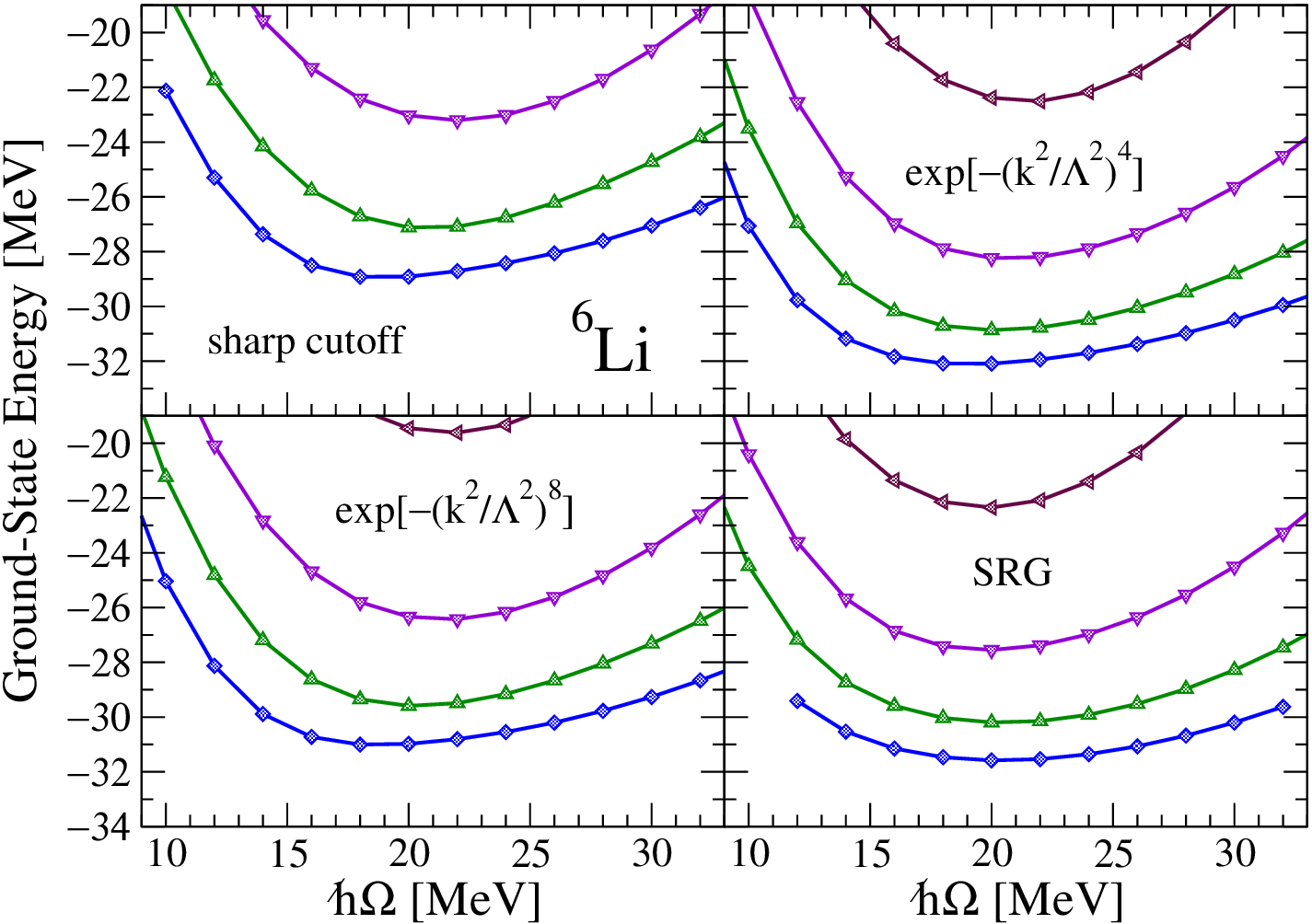}}
  \vspace*{-.1in}
  \caption{Ground-state  energy of \Lisix\ as a function of \hbaromega\
    for a sharp and two smooth $\vlowk$ potentials with $\Lambda = 2\infm$
    and an SRG potential with $\lambda = 2\infm$.
            The initial potential is the 500\,MeV \NthreeLO\ NN-only
	    potential from Ref.~\cite{N3LO}.  The legend from
            Fig.~\ref{fig:H2_srg_composite} applies here.}
  \label{fig:Li6_vlowk_composite}
\end{figure}

In previous comparisons between the SRG with flow parameter $\lambda$
and $\vlowk$ interactions with momentum cutoff $\Lambda$, 
the characteristics of the potential and its predictions 
at low energy were found to be strikingly
similar in two-body calculations when $\lambda \approx \Lambda$.  
The change in convergence 
with different $\Lambda$'s is similar to that shown for
different $\lambda$'s in Section~\ref{sec:srgenergies}. 
Ground-state energies for
$^2$H, $^3$H, $^4$He, and $^6$Li
are plotted against the harmonic-oscillator
parameter \hbaromega\ for a range of $\Nmax$ values  
in Figs.~\ref{fig:H2_vlowk_composite} through
\ref{fig:Li6_vlowk_composite}.  
Each figure shows results  
for a sharp and two smooth $\vlowk$ potentials with $\Lambda = 2\infm$
and an SRG potential with $\lambda = 2\infm$.
The initial potential is the 500\,MeV
N$^3$LO potential from Ref.~\cite{N3LO}.

We find that 
$\vlowk$ potentials with smoother cutoffs (but the same $\Lambda = 2\infm$)
converge more uniformly and that lower $\Nmax$ estimates are
superior.  
The latter advantage seems to largely disappear as the
space gets larger (e.g., see $^6$Li).  However, extrapolations may 
be more robust with the SRG and the smoother $\vlowk$ potentials. 
Note that the SRG \emph{is} a smooth cutoff low-momentum
potential.  With
$T_D$ defined with a single power of $T$, its characteristics
are closest
to those of a $\vlowk$ potential with exponential cutoff 
$\exp[-(k^2/\Lambda^2)^4]$ with $\lambda \approx \Lambda$.

The UCOM framework is an alternative approach
to using unitary transformations to soften
NN potentials and reduce correlations.  
We can compare the observed SRG convergence to that found for
the UCOM potential in Ref.~\cite{Roth:2005pd}, where results for
$^3$H and $^4$He starting from AV18 are shown in their Fig.~8. 
For $^3$H at $\Nmax = 6$ the UCOM result is about 2\,MeV from the converged
value and
at $\Nmax = 12$ the discrepancy is about 0.75\,MeV. 
The SRG calculation with $\lambda = 2\infm$ is 0.5\,MeV from
converged at $\Nmax = 6$ and only 25\,keV away at $\Nmax = 12$.
With $\lambda = 1.5\infm$, the discrepancies are 90 and 4\,keV, 
while with $\lambda = 3.0\infm$, they are 1.9\,MeV and 200\,keV.
If we consider $^4$He, the UCOM result at $\Nmax=6$
is about 2.5\,MeV and at $\Nmax = 10$ about 1\,MeV from converged.
For the SRG, the $\lambda = 1.5\infm$ discrepancies are 30 and 1\,keV
and
at $\lambda = 3.0\infm$ they are about 2.6 and 0.4\,MeV.
Therefore we conclude that, in terms of convergence, 
the UCOM potential from Ref.~\cite{Roth:2005pd} 
and the SRG $\lambda = 3.0\infm$ potential are roughly equivalent.

\section{Running of Ground-State Energies}
\label{sec:running}

The SRG flow equation guarantees a unitary transformed Hamiltonian and
exactly the same values for observables 
only if the entire Hamiltonian is kept.  However,
the commutators in Eq.~(\ref{eq:commutator}) generate many-body
interactions as $\lambda$ is lowered (e.g., substitute second-quantized
operators) and therefore in practice there will always be a truncation.  
But this is also true of the 
initial EFT
Hamiltonian, which will already have many-body interactions to all orders
that evolve with $\lambda$.  However, there is a chiral EFT
power-counting hierarchy that establishes natural sizes for the
contributions from many-body forces.  This enables truncation at few-body
interactions with a controlled error (at N$^3$LO this includes a long-range
four-body interaction).  Therefore the key to the usefulness
of the SRG for nuclear structure is maintaining this hierarchy with a
comparable truncation error. 
In the present calculations, keeping only two-body
interactions means that the transformations are only approximately
unitary  for $A\geqslant 3$ 
and few-body binding energies will change (``run'') with
$\lambda$. By tracking this running, we 
can determine the expected net 
contributions from many-body forces in a more complete calculation.
The useful range of $\lambda$ is
where the variations are of natural size (i.e.,
the same order as the EFT truncation errors).  

For the lower $\lambda$'s in the lighter nuclei, our predictions for
ground-state energies are fully converged.  However, in other cases we
need to extrapolate the energies to $\Nmax = \infty$.
If $\alpha$ labels the \hbaromega\ values and $i$ the
$\Nmax$ values for each $\alpha$, a possible
model for ground-state energies is
\beq
  E_{\alpha i} = E_\infty + A_\alpha\, e^{-b_\alpha N_i}
  \;,
  \label{eq:Ealphai}
\eeq
where $A_\alpha$ and $b_\alpha$ are constants.
The goal of a fit to calculations such as those in Section~\ref{sec:srgenergies}
is to determine the common parameter $E_\infty$, which
is the estimate for the ground-state energy extrapolated to $\Nmax = \infty$. 
Presumably this approach will 
work better with larger nuclei for which $\Nmax$ may become a good
logarithmic measure of the number of states.    

Given the model in Eq.~(\ref{eq:Ealphai}),
one way to proceed is to
minimize the function
\beq
   f(E_\infty,\{A_\alpha\},\{b_\alpha\})
     = \sum_{\alpha,i} (E_{\alpha i} - E_\infty - A_\alpha e^{-b_\alpha N_i})^2
            /\sigma_{\alpha i}^2
	    \;,
\eeq
where we have allowed the possibility for different weights
depending on \hbaromega\ and $\Nmax$.
As posed this is a nonlinear least-squares minimization problem with
many parameters, for which robust solutions are difficult to find.  
An alternative approach is to recast the problem so that all parameters 
except $E_\infty$
are treated as linear parameters.  This is possible because our problem
is strictly variational, so $E_{\alpha i} - E_\infty > 0$ for all
$\alpha$ and $i$, which means that the logarithm of Eq.~(\ref{eq:Ealphai})
is well defined.
So one considers instead
\beq
   \log(E_{\alpha i} - E_\infty) = 
     \log A_\alpha - b_\alpha N_i
     \equiv
     a_\alpha - b_\alpha N_i
     \; .
     \label{eq:logE}
\eeq
For each value of $E_\infty$, this is a linear 
least-squares problem in the $\{a_\alpha\}$
and $\{b_\alpha\}$ parameters.  
Thus we have a one-dimensional constrained
minimization problem with the function
\beq
  g(E_\infty) = \sum_{\alpha,i} 
    ( \log(E_{\alpha i} - E_\infty) - a_\alpha - b_\alpha N_i )^2
    /\sigma_{\alpha i}^2 
    \;,
    \label{eq:residual}
\eeq
where the $\{a_\alpha\}$
and $\{b_\alpha\}$ are determined directly within the function $g$ by invoking
a constrained linear least-squares minimization routine.
The constraint is the bound 
$E_\infty \leqslant \min(\{E_{\alpha i}\})$,
where $E_\infty < 0$ and ``min'' means ``most negative''.
Again, we allow for weights depending on $\Nmax$ and/or \hbaromega.

\begin{figure}[tp] 
 \centerline{\includegraphics*[width=4.in]{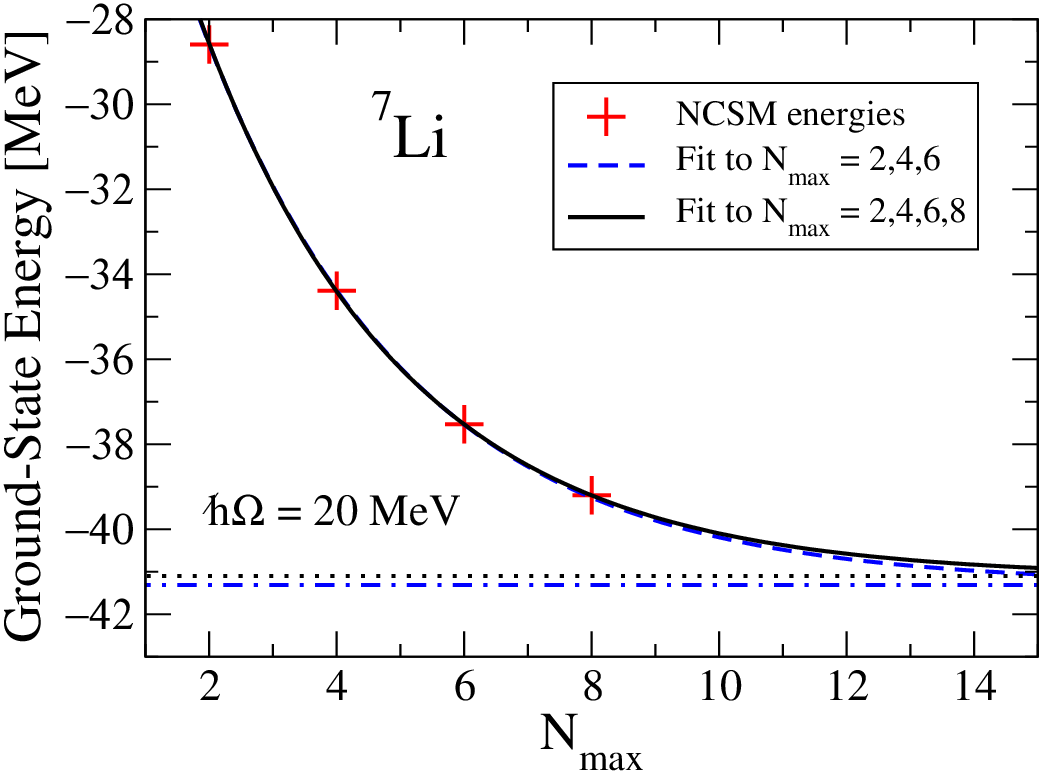}}
 \vspace*{-.1in}
 \caption{Fits to the NCSM ground-state energies for $^7$Li at $\lambda = 2\infm$
 with \hbaromega\ = 20\,MeV  
   according to the model in Eq.~(\ref{eq:Ealphai}), using the procedure
   described in the text.  The dashed line uses the NCSM energies 
   (pluses) up to $\Nmax = 6$,
   with the extrapolated value of $E_\infty$ shown as the dot-dash line, while
   the solid line uses up to $\Nmax = 8$, with the dotted line the
   extrapolated value of $E_\infty$.}
 \label{fig:extrapolate}
 
 \vspace*{.2in}
\end{figure}

In the present investigation, we apply Eq.~(\ref{eq:residual}) with
only the \hbaromega\ value that yields the lowest energy in the largest
space, weighting different $\Nmax$ by the slope of the energy vs.\ $\Nmax$,
and using the spread of 
results from neighboring \hbaromega\ values to determine
a conservative confidence interval for the extrapolation.
This procedure is applied for all results in this section. An example
of the fit to NCSM data is given in Fig.~\ref{fig:extrapolate}.
A more complete and systematic study of extrapolation in the NCSM will be
given in Ref.~\cite{Pieter}.

\newcommand{\phansp}{\phantom{$0$}}
\setlength{\tabcolsep}{4.5pt}

\begin{table}
\caption{Ground-state energies and optimal \hbaromega\ (in MeV) of light nuclei 
            for SRG-evolved potentials.
	    When an uncertainty is given it represents a confidence
            interval derived from an extrapolation to $\Nmax = \infty$.
            Otherwise the energy is converged to the digits shown.
            }
\label{tab:energies}
\begin{tabular}{|c|cc|cc|cc|cc|cc|}
 \hline
 \multicolumn{1}{|c}{} & \multicolumn{2}{|c|}{$^3$H} & \multicolumn{2}{c|}{$^4$He}
 & \multicolumn{2}{|c|}{$^6$He} 
   & \multicolumn{2}{c|}{$^6$Li}  & \multicolumn{2}{|c|}{$^7$Li}\\[0pt]
  $\lambda$  & \hbaromega & $E_{\rm gs}$  
         & \hbaromega & $E_{\rm gs}$ & \hbaromega & $E_{\rm gs}$  & \hbaromega & $E_{\rm gs}$
   & \hbaromega & $E_{\rm gs}$\\[0pt]
 \hline 
  $\infty$ & -- & $-7.85$          &  42 & $-26.1(8)$ &  &   &  & &  & \\[0pt]
  $3.0$\phansp & 28 & $-8.29$ &  34 & $-27.5(3)$ & 28 & $-28(1)$  & 28 & $-31.5(8)$ & 24 & $-38.7(30)$        \\[0pt]
  $2.5$\phansp & 24 & $-8.41$ &  28 & $-28.2(2)$ & 24 & $-28.9(3)$ & 24 & $-32.1(3)$ & 24 & $-38.7(20)$        \\[0pt]
  $2.25$ & 22 & $-8.47$       &  24 & $-28.6(1)$ & 22 & $-29.4(2)$	 & 22 & $-32.5(2)$ & 22 & $-40.3(10)$  \\[0pt]
  $2.0$\phansp & 18 & $-8.53$ &  24 & $-28.90$   & 20 & $-30.0(1)$ & 20 & $-33.1(1)$ & 20 & $-41.2(5)$        \\[0pt]
  $1.75$ & 16 & $-8.55$       &  20 & $-29.13$   & 16 & $-30.6$ 	 & 18 & $-33.6$    & 18 & $-41.7(4)$  \\[0pt]
  $1.5$\phansp & 12 & $-8.48$ &  18 & $-28.86$   & 14 & $-30.7$    & 16 & $-33.7$    & 16 & $-42.0(3)$        \\[0pt]
  $1.25$ & 10 & $-8.21$       &  14 & $-27.58$   & 12 & $-29.9$ 	 & 12 & $-32.9$    & 12 & $-41.1(2)$  \\[0pt]
  $1.0$\phansp & 8 & $-7.63$  &  14 & $-24.80$   & 10 & $-27.4$    & 10 & $-30.4$    & 12 & $-37.8(2)$        \\[0pt]
 \hline
\end{tabular}
\end{table}

The extrapolated results for ground-state energies (in MeV) are summarized
in Table~\ref{tab:energies} along with the value of \hbaromega\ (in MeV) for which
the energy is minimal in the largest spaces.   
The uncertainties represent
conservative confidence intervals.  When not given, the result should
be accurate to the digits displayed.
The $^3$H and $^4$He
results are fully converged for $\lambda \leqslant 2\infm$.
For
larger $\lambda$ values the $^3$H energies are evaluated in a large basis
($\Nmax = 48$) to ensure convergence for all $\lambda$
and the $^4$He results are extrapolated.
All of the results for $A > 4$ are extrapolated to some degree, 
with increasingly
small extrapolations
with decreasing $\lambda$ (and energies for the bare potentials could
not be reliably extrapolated in the spaces used).

\begin{figure}[p]

 \centerline{\includegraphics*[width=3.6in]{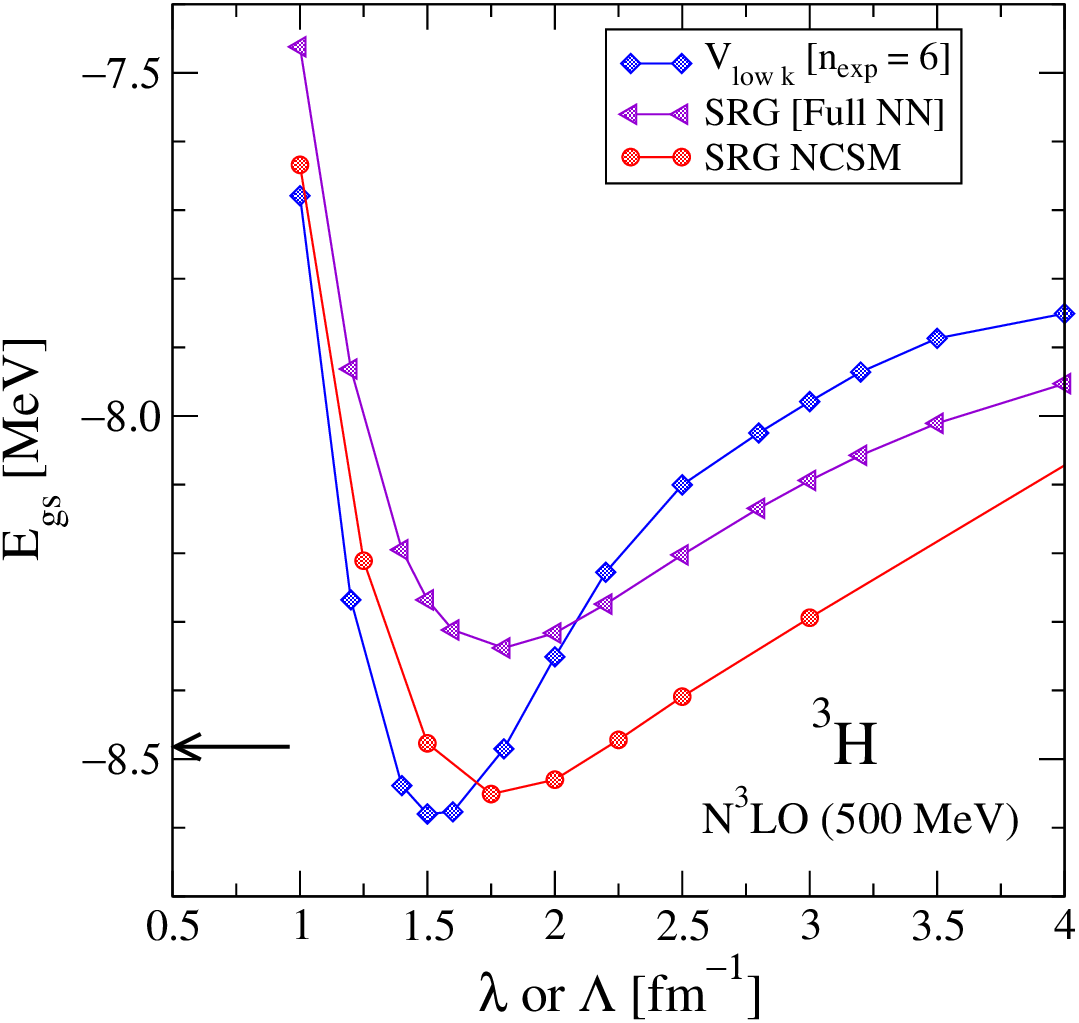}}
 \vspace*{-.1in}
 \caption{Plot of the ground-state energy of the triton
 vs.\ $\lambda$ (if SRG)
 or $\Lambda$ (if $\vlowk$) for potentials evolved 
 from the  500\,MeV
   N$^3$LO NN-only potential from Ref.~\cite{N3LO}.  
   The arrow marks the experimental binding.
   The two SRG curves show the difference between evolving the
   full NN interaction and evolving the neutron-proton interaction only.}
 \label{fig:3bodyH3}
 
 \vspace*{.1in}
 
 \centerline{\includegraphics*[width=3.0in]{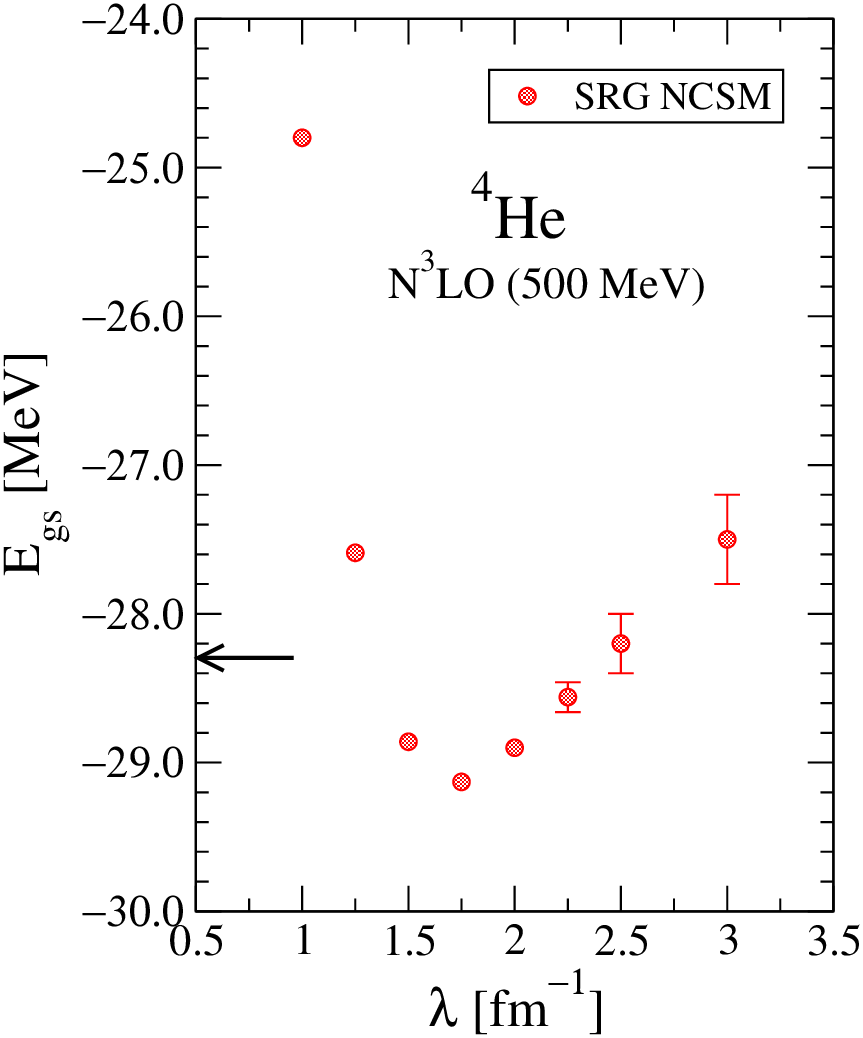}
             \includegraphics*[width=3.0in]{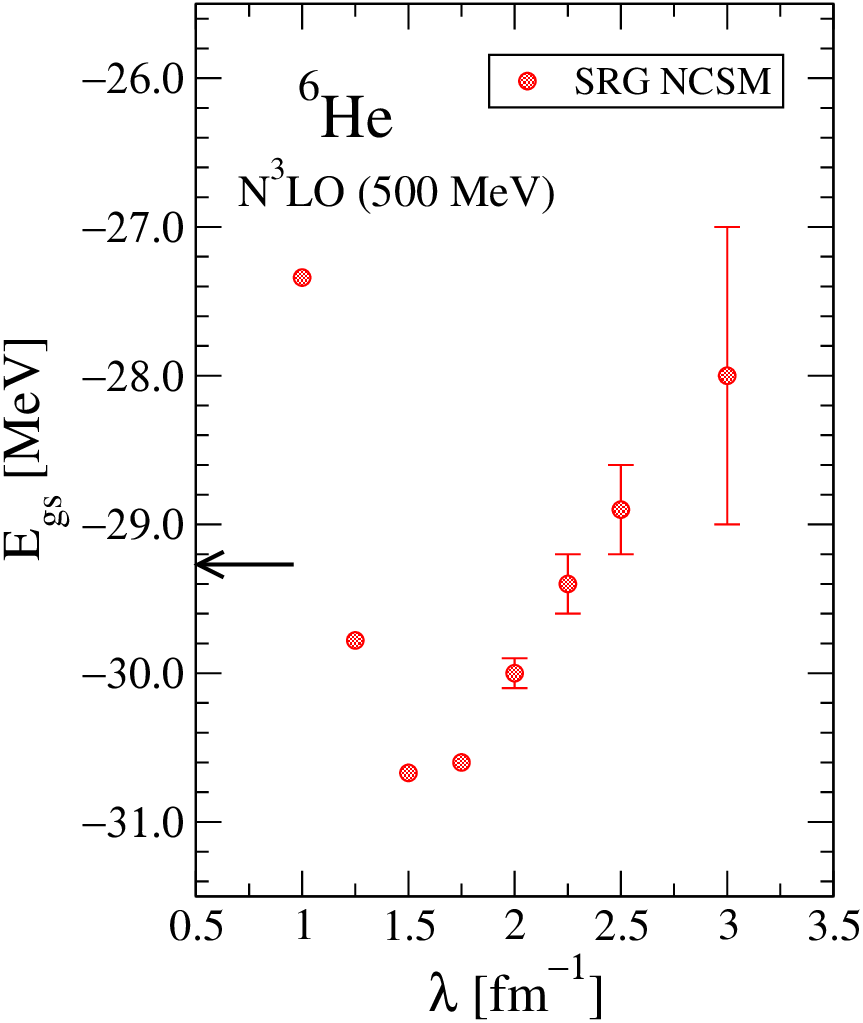}
 }
 \vspace*{-.1in}
 \caption{Plot of the ground-state energy of  
   $^4$He and $^6$He vs.\ $\lambda$  for potentials 
    evolved by the SRG from the  500\,MeV
   N$^3$LO NN-only potential from Ref.~\cite{N3LO}.
   Conservative error bars have been included with
   the larger $\lambda$'s, for which an extrapolation is needed.    
   The arrow marks the experimental binding.}
 \label{fig:3bodyHe4}
   
\end{figure}

\begin{figure}[p]
 
 \centerline{\includegraphics*[width=3.0in]{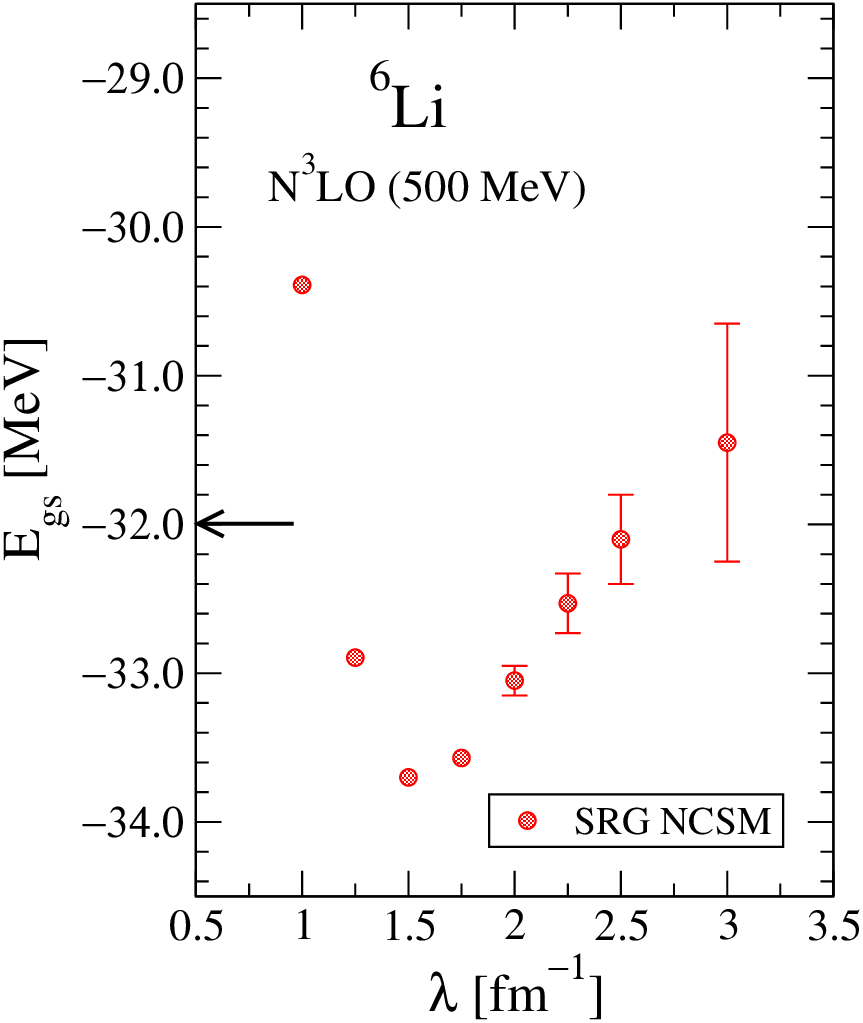}
             \includegraphics*[width=3.0in]{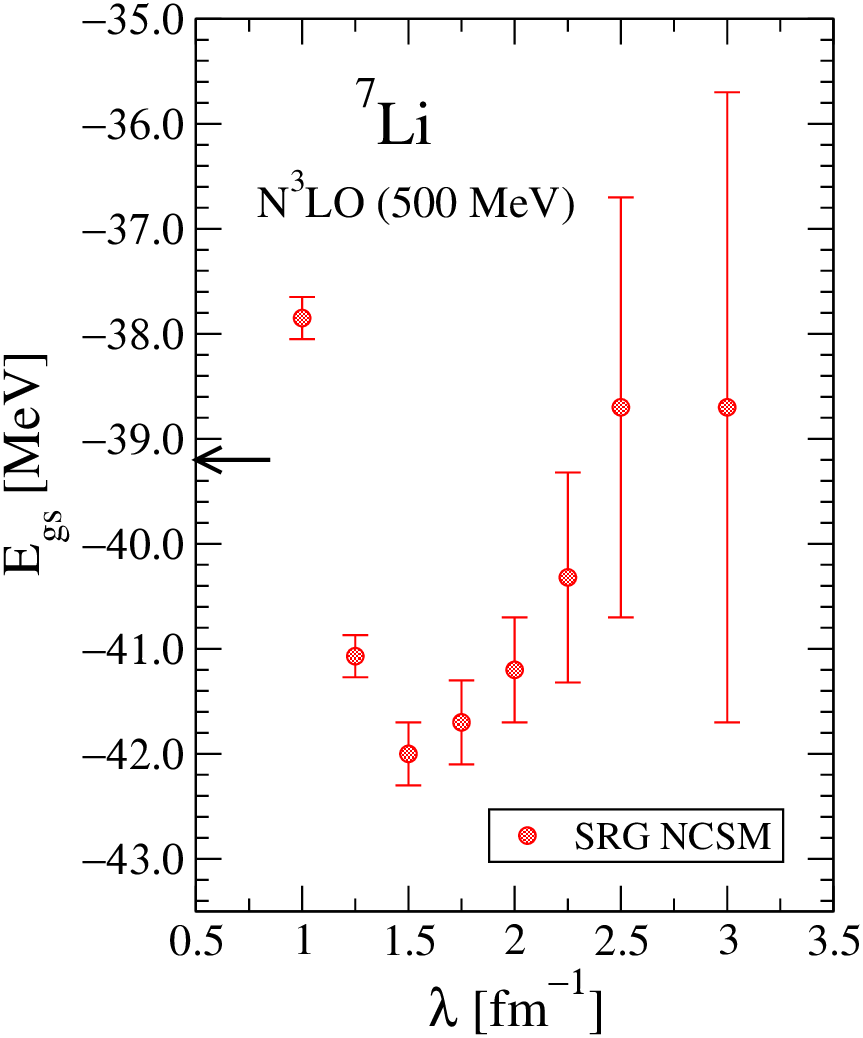}
 }
 \vspace*{-.1in}
 \caption{Plot of the ground-state energy  of $^6$Li and $^7$Li
 vs.\ $\lambda$  for SRG potentials
   evolved from the  500\,MeV
   N$^3$LO NN-only potential from Ref.~\cite{N3LO}.  
   Conservative error bars have been included with
   the larger $\lambda$'s, for which an extrapolation is needed.    
   The arrow marks the experimental binding.}
 \label{fig:3bodyLi6}
 
 \vspace*{.1in}
 
 \centerline{\includegraphics*[width=3.7in]{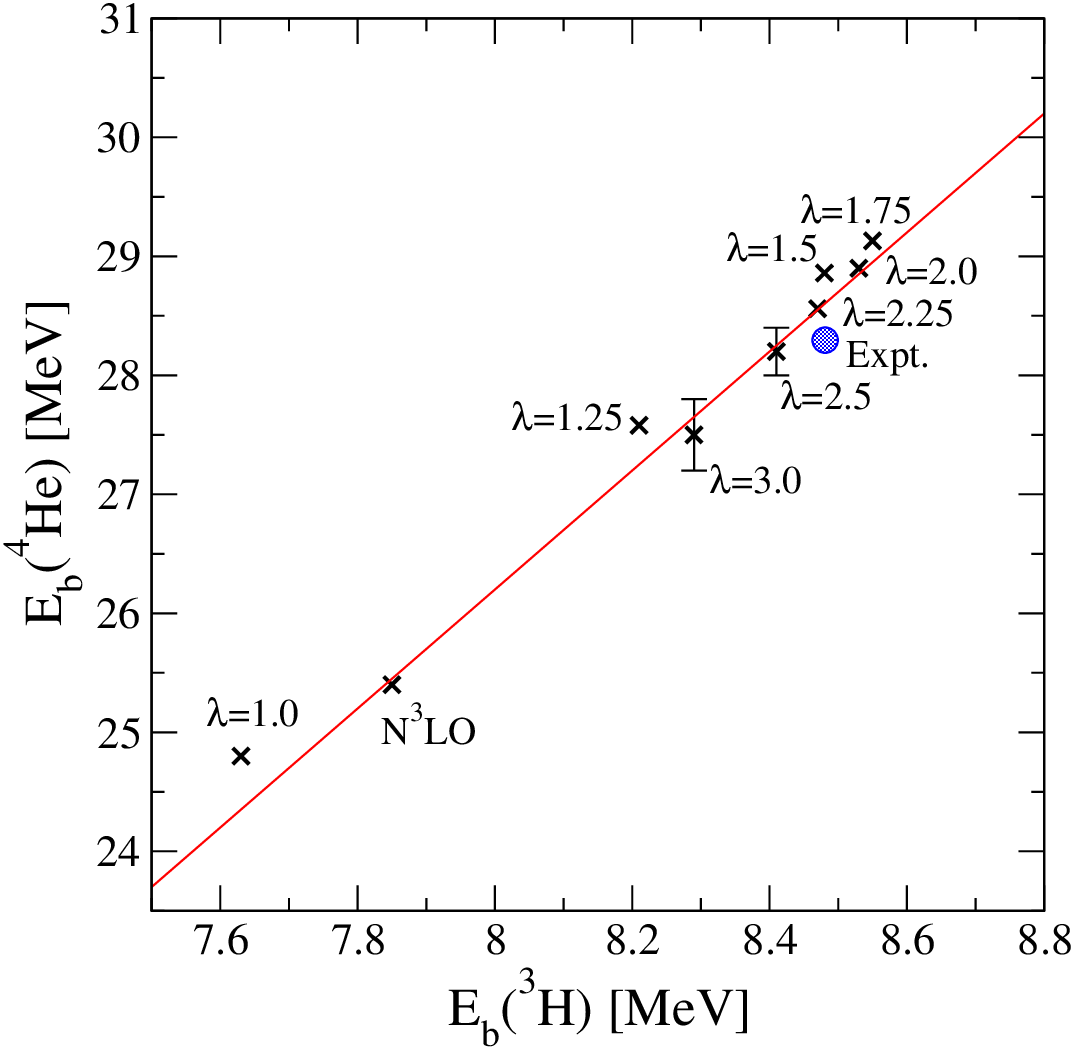}}
 \vspace*{-.1in}
 \caption{Tjon line for SRG potentials 
    evolved from the  500\,MeV
   N$^3$LO NN-only potential from Ref.~\cite{N3LO}
   (with $\lambda$ in fm$^{-1}$).  
   Conservative error bars have been included with
   the larger $\lambda$'s, for which an extrapolation is needed. 
   The result for $^4$He with the bare potential is from
   Ref.~\cite{Navratil04}. 
   The line is a fit to NN potentials from
   Ref.~\cite{Nogga:2000uu}.
 }
 \label{fig:3bodyTjon}
   
\end{figure}

\begin{figure}[p]
 
 \centerline{\includegraphics*[width=3.7in]{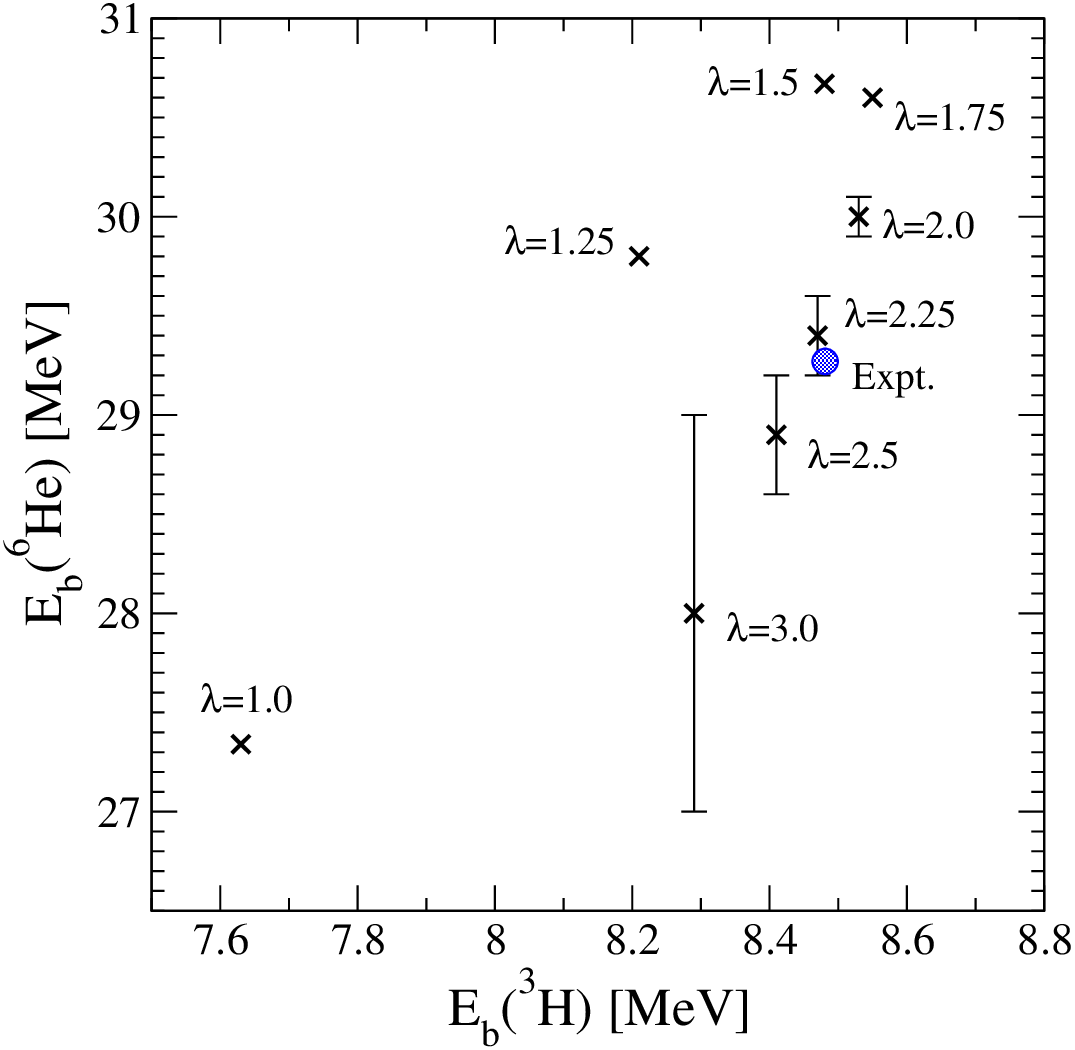}}
 \vspace*{-.1in}
 \caption{Binding energy correlation plot between $^6$He and $^3$H 
 for SRG potentials evolved from the  500\,MeV
   N$^3$LO NN-only potential from Ref.~\cite{N3LO} 
   (with $\lambda$ in fm$^{-1}$).   
   Conservative error bars have been included with
   the larger $\lambda$'s, for which an extrapolation is needed.  }
 \label{fig:3bodyTjonHe6}
 
 \vspace*{.2in}
 
 \centerline{\includegraphics*[width=3.7in]{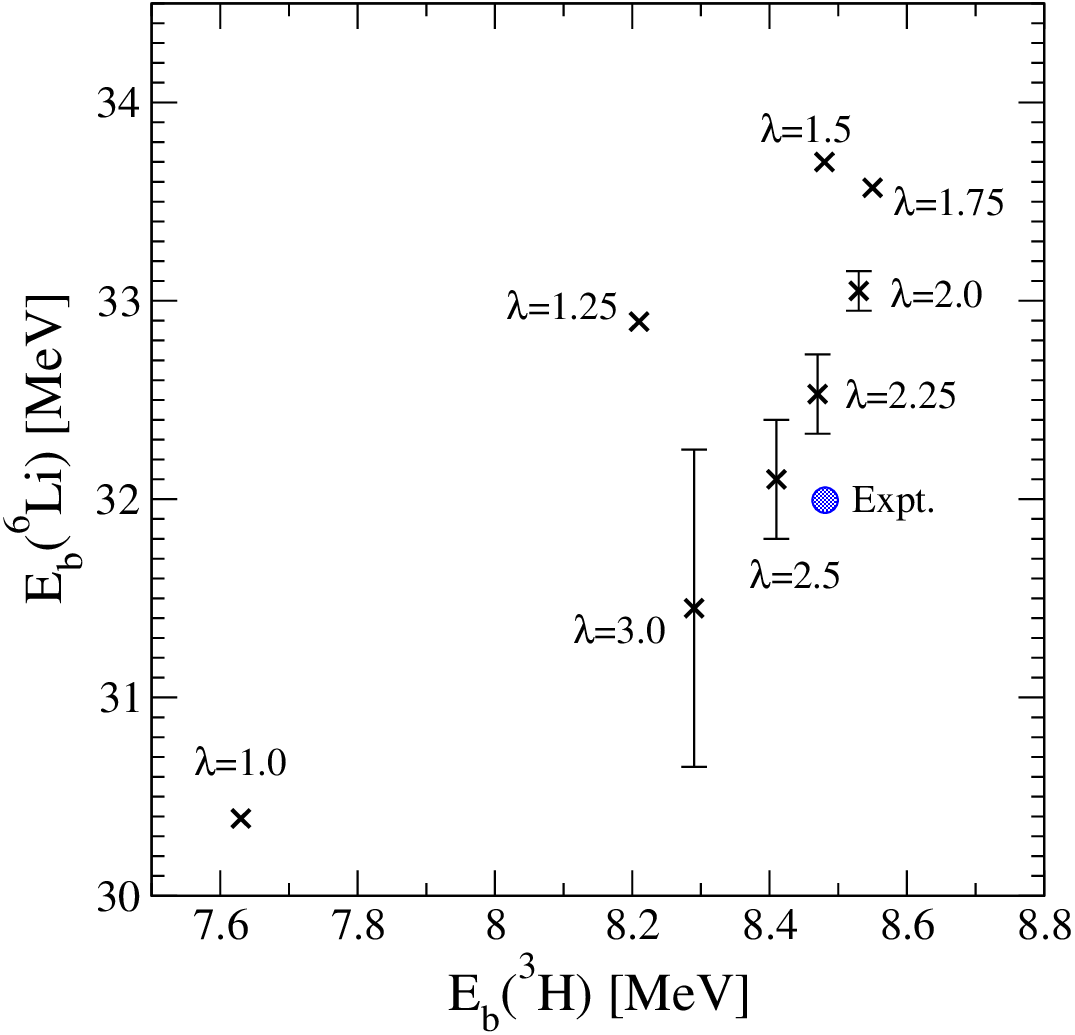}}
 \vspace*{-.1in}
 \caption{Binding energy correlation plot between $^6$Li and $^3$H 
 for SRG potentials evolved from the  500\,MeV
   N$^3$LO NN-only potential from Ref.~\cite{N3LO}
   (with $\lambda$ in fm$^{-1}$).  
   Conservative error bars have been included with
   the larger $\lambda$'s, for which an extrapolation is needed.  }
 \label{fig:3bodyTjonLi6}
   
\end{figure}

\begin{figure}[tp]
 
 \centerline{\includegraphics*[width=4.in]{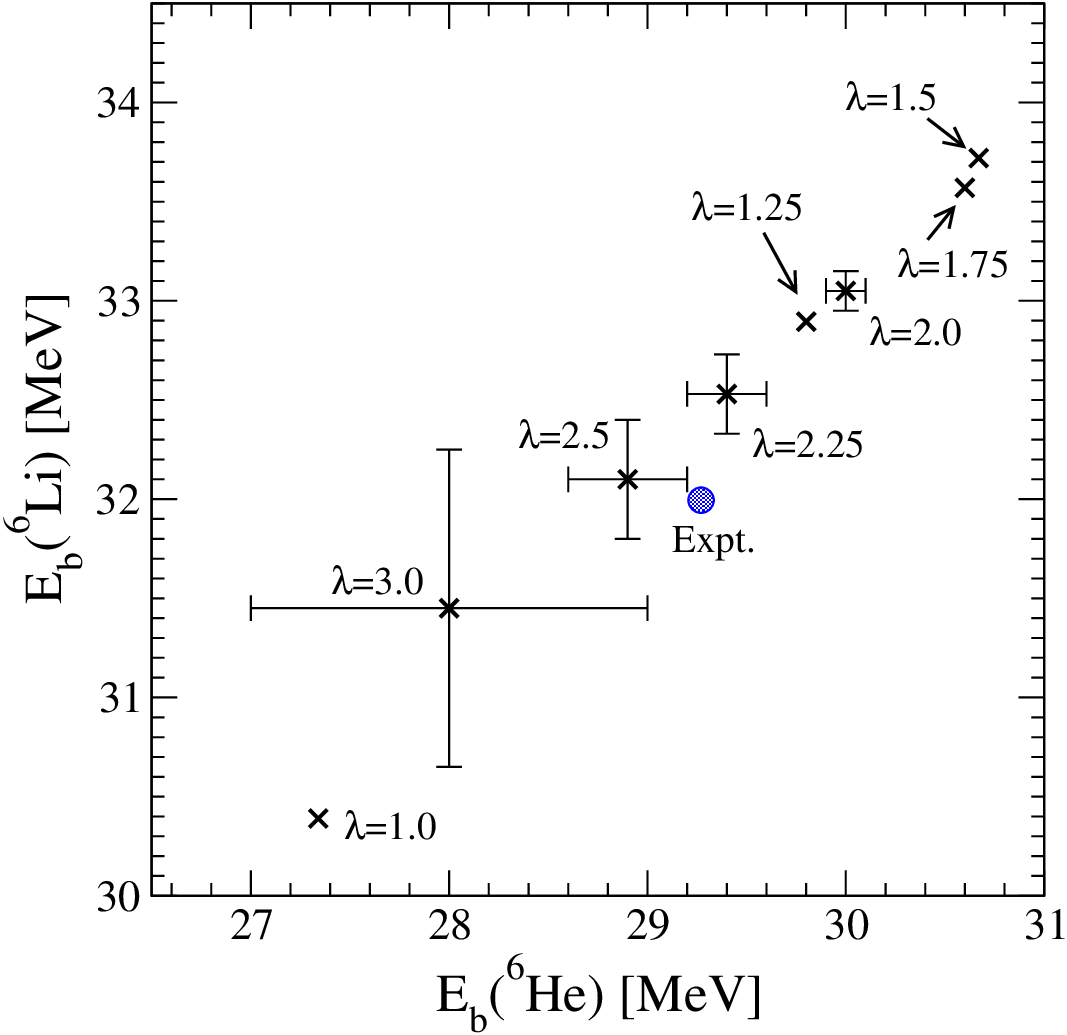}}
 \vspace*{-.1in}
 \caption{Binding energy correlation plot between $^6$Li and $^6$He 
 for SRG potentials evolved from the  500\,MeV
   N$^3$LO NN-only potential from Ref.~\cite{N3LO}
   (with $\lambda$ in fm$^{-1}$).  
   Conservative error bars have been included with
   the larger $\lambda$'s, for which an extrapolation is needed.  }
 \label{fig:3bodyTjonLi6He6}
 
 \vspace*{.2in}
\end{figure}

If we calculate the energy of a nucleus using the SRG-evolved NN 
interaction only, the running of the energy is the running of the
\emph{net} three-body (and higher-body) contribution to the energy.
Over a wide range of cutoffs,
many-body interactions generated in the RG will be the same scale
as the starting  3N interactions in the chiral EFT.
Figures~\ref{fig:3bodyH3} to \ref{fig:3bodyLi6} show the NN-only running 
for $^3$H, $^4$He, $^6$He,
and $^6$Li.
The energy plotted is either the converged value or
extrapolated from the best \hbaromega\ with an error bar (i.e., 
a confidence interval) as described above.  
Figures~\ref{fig:3bodyTjon} to \ref{fig:3bodyTjonLi6He6}
are binding-energy correlation plots between various nuclei.   

The pattern of SRG running 
for $^4$He, $^6$He, $^6$Li, and $^7$Li is qualitatively
similar to the pattern for $^3$H shown in Fig.~\ref{fig:3bodyH3}.
That is, a slow increase in the binding energy as $\lambda$ decreases
to $\lambda = 1.5\mbox{--}1.8\infm$ and then a steep decrease.
The two SRG curves in Fig.~\ref{fig:3bodyH3} show the difference
between evolving the neutron-proton interaction only and evolving
the full NN interaction.
The SRG pattern is also qualitatively similar
to the running for $\vlowk$ potentials~\cite{Vlowk3N,Bogner:2006vp}.
As observed for $\vlowk$~\cite{Vlowk3N,BognerNucmatt}, the size of
three-body contributions is natural according to chiral
EFT power counting.
This is true for the net energy within the entire range of $\lambda$
shown in the figures, 
but future calculations are needed to investigate whether
the individual contributions from long-range
and short-range 3N forces remain natural. 
The smoothness of the running adds credibility to the
extrapolations and the associated estimates of confidence intervals;  
when the estimate has been improved (e.g., from a larger $\Nmax$ result), 
the running and correlation plots
have become more regular.

The correlation plot for $^4$He vs.\ $^3$H shows the expected
movement along the Tjon line, with a slight loop closing to the
left as $\lambda$ decreases.  The latter behavior is  amplified
in the $^6$He vs.\ $^3$H and $^6$Li vs.\
$^3$H plots.  The plot of $^6$Li vs.\ $^6$He is very close to linear.
Tjon lines are also found in coupled cluster calculations of
$^{15}$O, $^{16}$O, $^{17}$O, $^{15}$N, and $^{17}$F~\cite{Gaute}. 
From these plots one sees some $\lambda$'s for which predicted ground-state
energies are quite close to experiment (e.g., $\lambda = 2.25\infm$).
These may be good choices for calculations for which small net three-body
contributions are desirable.  However, we emphasize that this is not a global
trend.  Indeed, nuclear matter does not saturate with only NN 
interactions at such
$\lambda$'s (except perhaps at very large density).  Thus calculations
in heavier nuclei (e.g., $^{16}$O, $^{40}$Ca, \ldots) will be increasingly
overbound.

\section{Other Observables}
  \label{sec:observables}
  
In this section, we show some limited results of excitation energies
and radii for $^7$Li.  Because we expect sensitivity to three-body
contributions, we include these results primarily to assess convergence
trends as a baseline for future 3N calculations
and to compare to previous results from Ref.~\cite{Nogga:2005hp}.  

\begin{figure}[pth]
  \centerline{\includegraphics*[width=3.95in,angle=0]%
{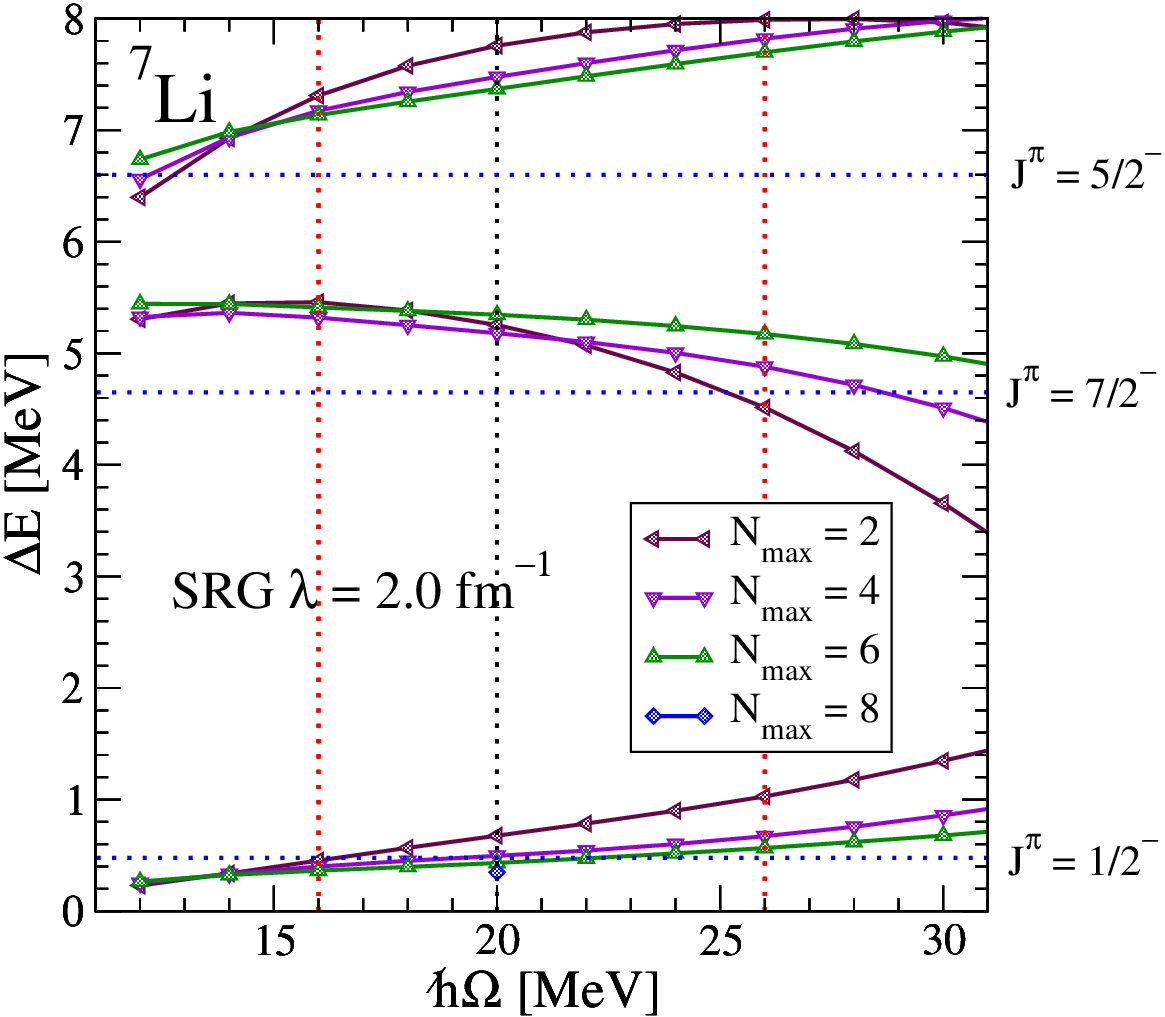}}
  \vspace*{-.1in}
  \caption{Excitation energies of the lowest natural-partiy
  states of \Liseven\ 
    as a function of \hbaromega\
    for $\lambda = 2.0\infm$.
            The initial potential is the 500\,MeV \NthreeLO\ NN-only
	    potential from Ref.~\cite{N3LO}.
            The horizontal dotted lines are the experimental values
            while the vertical dotted lines mark the optimal \hbaromega\
            value for the ground-state energy (middle) and the range for
            which estimates are close to this. }
  \label{fig:Li7ex_srg_composite}

 \vspace*{.08in}
 
  \centerline{\includegraphics*[width=3.95in,angle=0]%
{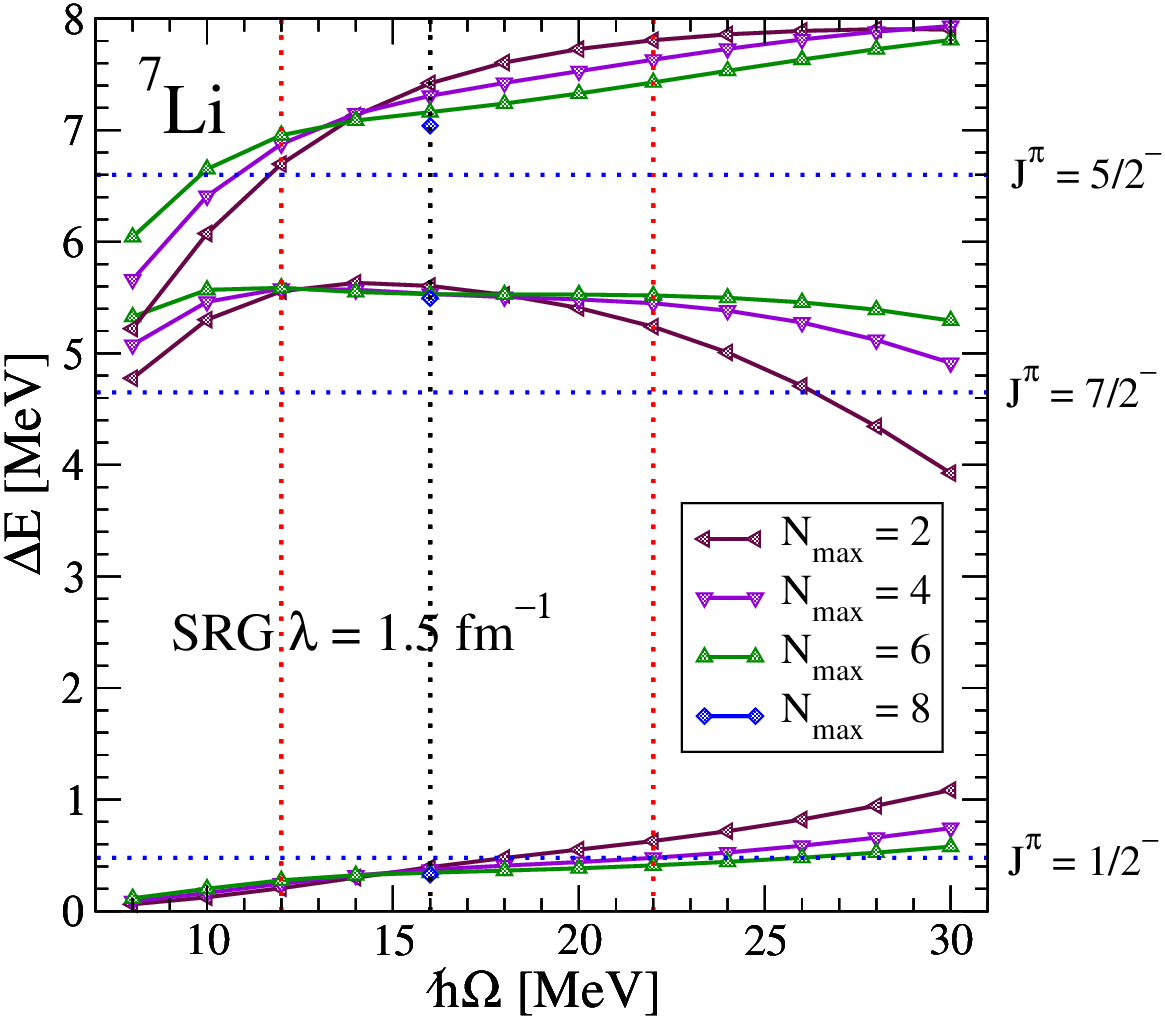}}
  \vspace*{-.1in}
  \caption{Excitation energies of the lowest natural-parity states of \Liseven\ 
    as a function of \hbaromega\
    for $\lambda = 1.5\infm$.
            The initial potential is the 500\,MeV \NthreeLO\ NN-only
	    potential from Ref.~\cite{N3LO}.
            The horizontal dotted lines are the experimental values
            while the vertical dotted lines mark the optimal \hbaromega\
            value for the ground-state energy (middle) and the range for
            which estimates are close to this. }
  \label{fig:Li7ex_srg_composite2}
\end{figure} 

The spectrum of low-lying natural-parity
excited states in \Liseven\ is shown in 
Figs.~\ref{fig:Li7ex_srg_composite} and
\ref{fig:Li7ex_srg_composite2} for $\lambda = 2.0\infm$ and $1.5\infm$.  
The excitation energy is plotted against \hbaromega\ for different $\Nmax$
values to facilitate comparison to
similar plots from NCSM Lee-Suzuki calculations based on the same
potential~\cite{Nogga:2005hp}.
The vertical dotted lines mark the range of \hbaromega\ that has the lowest
ground-state energy estimates, 
with the middle line marking
the best energy estimate for the ground state.
The higher $\Nmax$ predictions are flatter as a function of \hbaromega,
as expected, and $\lambda = 1.5\infm$ converges faster than
$\lambda = 2.0\infm$.
However, we observe that the curves
are not as flat as in the corresponding plot
in Ref.~\cite{Nogga:2005hp}.
The spread from $\Nmax = 2$ to $\Nmax = 6$ is not large, but
also appears larger than in  Ref.~\cite{Nogga:2005hp}.

\begin{figure}[pth]
  \centerline{\includegraphics*[width=4.75in,angle=0]%
             {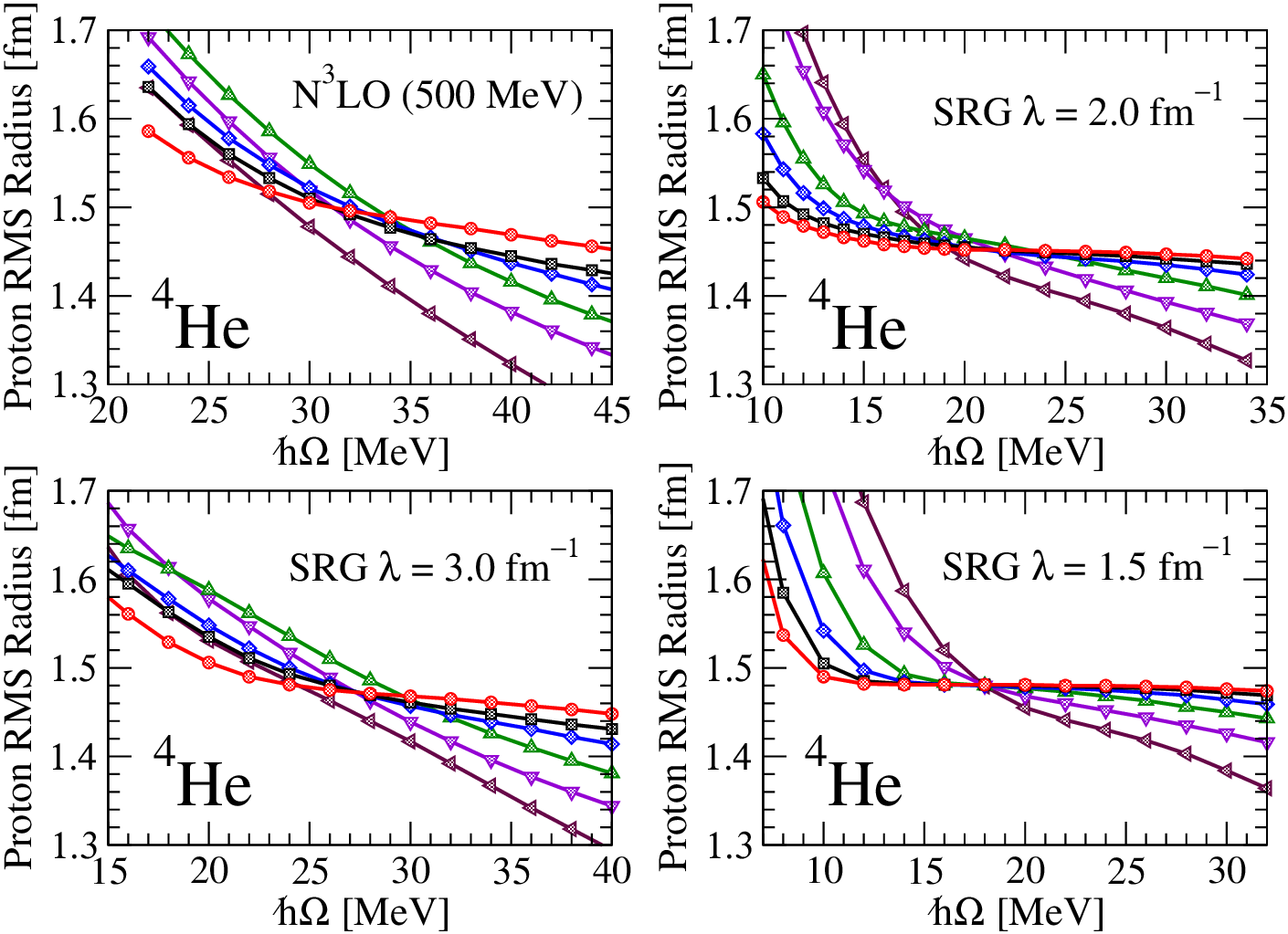}}
  \vspace*{-.1in}
  \caption{Point proton radii of \Hefour\ as a function of \hbaromega\
    at four different values of $\lambda$ ($\infty$, 3, 2, $1.5\infm$).
            The initial potential is the 500\,MeV \NthreeLO\ NN-only
	    potential from Ref.~\cite{N3LO}.  The legend from
            Fig.~\ref{fig:H2_srg_composite} applies here.}
  \label{fig:He4_rms_p_srg_composite}
  \vspace*{.1in}
  
  \centerline{\includegraphics*[width=4.75in,angle=0]%
             {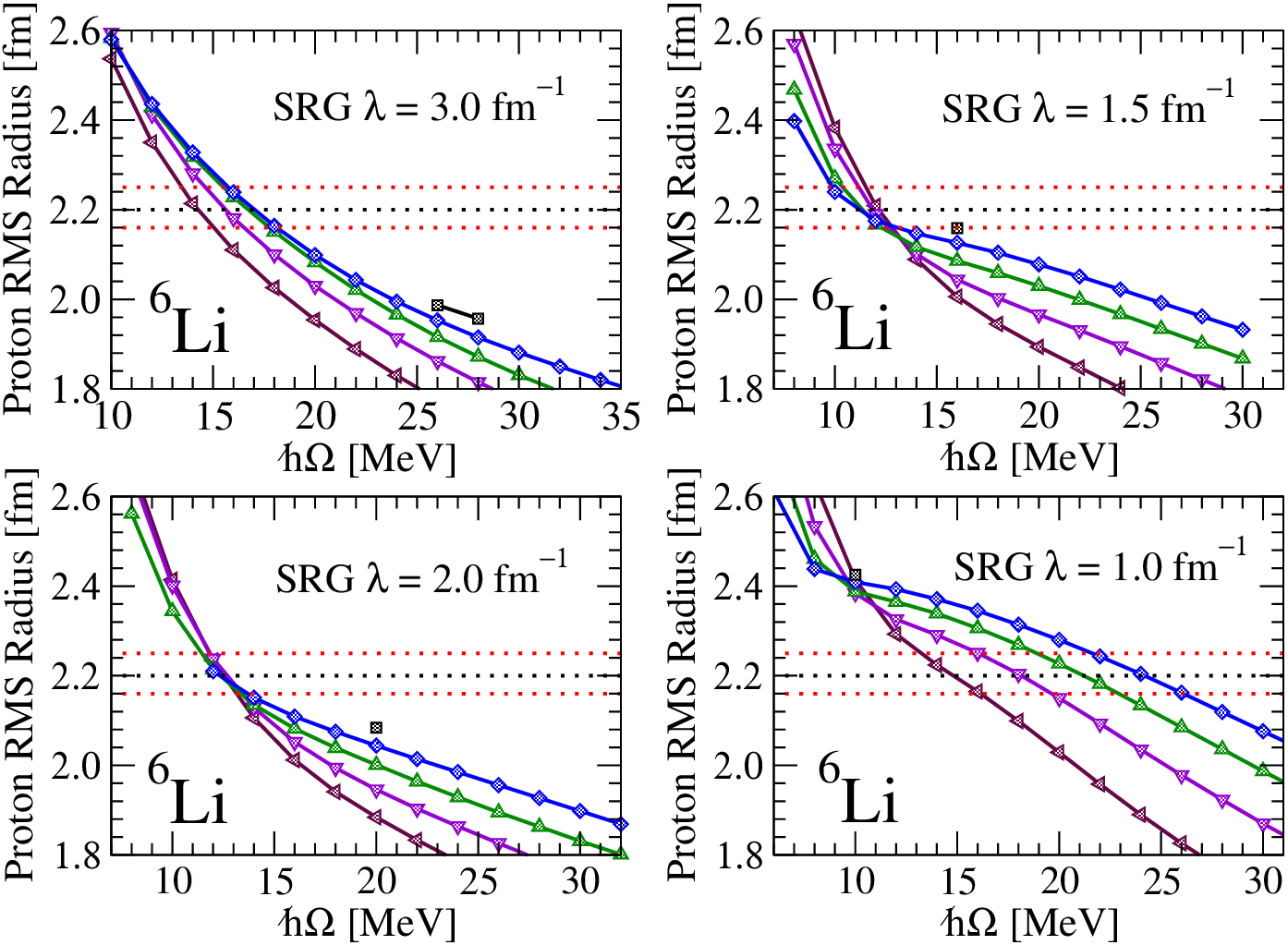}}
  \vspace*{-.1in}
  \caption{Point proton radii of \Lisix\ as a function of \hbaromega\
    at four different values of $\lambda$ (3, 2, 1.5, $1\infm$).
            The initial potential is the 500\,MeV \NthreeLO\ NN-only
	    potential from Ref.~\cite{N3LO}.  The three dotted lines
            are radii from   NCSM Lee-Suzuki
            calculations using the bare N$^3$LO potential 
            in larger spaces~\cite{Nogga:2005hp}.  
            The middle (black)
            line is without 3N
            interactions and the others are with two different 3N
            force fits~\cite{Nogga:2005hp}.
            The legend from Fig.~\ref{fig:Li6_srg_composite} applies
            here.%
            }
  \label{fig:Li6_rms_p_srg_composite}
  \vspace*{.1in}
\end{figure}

In Figs.~\ref{fig:He4_rms_p_srg_composite} 
and \ref{fig:Li6_rms_p_srg_composite}, we plot the point proton
radii of \Hefour\ and \Lisix\ as a function of \hbaromega, with
the different $\Nmax$ curves having the same legends as
in Figs.~\ref{fig:H2_srg_composite} and \ref{fig:Li6_srg_composite}.
These plots are made using the bare operator for $r^2$ at all $\lambda$,
rather than an evolved operator.
(Note that $r^2$ for the proton is not an observable.) 
Since $r^2$ samples primarily long distances, its
matrix elements depend weakly on the cutoff~\cite{Bogner:2006vp}.  
For the deuteron, the rms radius changes from
the $\lambda = \infty$ value by less than 1\% by $\lambda = 1.5\infm$
and by about 4\% by $\lambda = 1.0\infm$.
Our observations are consistent with the investigations reported 
recently on the range-dependence of another renormalization 
scheme~\cite{Stetcu}.

For $^4$He, we observe that 
with decreasing $\lambda$ the curves for a given $\Nmax$
get flatter; for $\lambda = 1.5\infm$ our $\Nmax = 12$ results are
essentially independent of \hbaromega.
The different $\Nmax$ curves cross each other in a
decreasing range of \hbaromega\ as $\lambda$ decreases.
For $\lambda = 1.5\infm$, \emph{all} of the curves for both
nuclei intersect at the same point.  If we consider the intersection
of the three highest $\Nmax$ results, it is close to
(but generally slightly below) 
the \hbaromega\ for the lowest ground-state energy estimate.  A similar
observation has been reported for the NCSM Lee-Suzuki calculations
in Ref.~\cite{Nogga:2005hp}.
The trend is consistent
with the idea that  if you have
the optimal \hbaromega\ for a given nucleus (which scales
with $A^{-1/3}$ and inversely with the average radius squared),
then the radius should be minimally sensitive to changes in the basis size.
For $^6$Li, the intersection point is well below the optimal \hbaromega\
for the energy except at very low cutoffs; the radii for larger
$\lambda$ are not converged in the largest spaces considered here.   

In Fig.~\ref{fig:Li6_rms_p_srg_composite}, the dotted lines
indicate the radii from the NCSM Lee-Suzuki calculations
of the bare N$^3$LO potential without
3N interactions (middle) and with two different
3N fits (top and bottom)~\cite{Nogga:2005hp}.  
The spread of these predictions
can be taken as an estimate of the expected 3N
contribution for various $\lambda$'s. 
The \hbaromega\
dependence here is steep, but if we take the intersection
point to determine the NN radius, then the results
for $\lambda = 3,2,1.5\infm$ are all compatible
with the radii from Ref.~\cite{Nogga:2005hp}.  
Moreover, the trend in the radius with decreasing $\lambda$ parallels
the running of the energy with $\lambda$ (see Fig.~\ref{fig:3bodyLi6}).

\section{Summary}
 \label{sec:summary}

We have performed
NCSM calculations that retain the variational principle
using RG-evolved potentials in momentum representation,
focusing on the SRG approach that
has recently been applied to nuclear structure physics.
While we have not performed 
an exhaustive test of alternative potentials, we have found no sign of
any special dependence on the initial potential.
The convergence of ground-state energies improves rapidly with 
decreasing $\lambda$ and with no unnatural contributions
from omitted three-body forces for these light nuclei.
Furthermore, all methods of evolution to low momentum give similar
convergence improvements, with smoother behavior associated with
smoother cutoffs.

With these calculations
we have set the stage for including 3N interactions~\cite{SRG3body}.
Just as varying $\lambda$ in Figs.~\ref{fig:3bodyH3} to \ref{fig:3bodyLi6} 
shows the scale of omitted
many-body contributions, similar figures when 3N is included will quantify
the net impact of higher-order interactions.
In the short term, three-body forces will be included in the
form of  N$^2$LO chiral 3N interactions, with the parameters fitted
(see Ref.~\cite{Vlowk3N} for motivation).
The development of evolved 3N potentials is proceeding in
parallel and should be available in the near future.
Once we have consistent 3N interactions, we will be able
to establish lower bounds to theoretical error bars for light nuclei
by calculating with different N$^3$LO cutoffs and assess the
impact of uncertainties in
the input chiral 3N interactions.

\begin{ack}
We thank E. Anderson, E. Jurgenson, A. Nogga, S. Ramanan and the referee
for useful comments.
This work was supported in part by the National Science 
Foundation under Grant Nos.~PHY--0354916 and PHY--0653312,  
the Department of Energy under Grant No.\
DE-FG02-87ER40371, 
the UNEDF SciDAC Collaboration under DOE Grant 
DE-FC02-07ER41457,
and the Natural Sciences and Engineering Research Council of Canada 
(NSERC). TRIUMF receives federal funding via a contribution agreement 
through the National Research Council of Canada.
\end{ack}


\end{document}